\documentclass{emulateapj}

\usepackage{natbib}
\newcommand{\tdyn   }{t_\mathrm{dyn    }}
\newcommand{\tcool  }{t_\mathrm{cool   }}
\newcommand{\tcoolff}{t_\mathrm{cool,ff}}
\newcommand{\tdiff  }{t_\mathrm{diff   }}
\newcommand{\texp   }{t_\mathrm{exp    }}
\newcommand{\timp   }{t_\mathrm{imp    }}
\newcommand{\tpeak  }{t_p               }

\newcommand{\rhocsm}{\rho_\mathrm{CSM}}
\newcommand{\rhoej }{\rho_\mathrm{ej }}

\newcommand{\CSM}{\mathrm{CSM}}
\newcommand{\dubs}{\mathrm{ss}}

\newcommand{\vejmax}{v_\mathrm{ej,max}}

\newcommand{\vejobs}{v_\mathrm{ej,o}}

\newcommand{\ugas  }{u_\mathrm{gas}}
\newcommand{\jff   }{j_\mathrm{ff}}
\newcommand{\fNT   }{f_\mathrm{NT}}
\newcommand{\neNT  }{n_{e,\mathrm{NT}}}
\newcommand{\gammin}{\gamma_\mathrm{min}}
\newcommand{\epse  }{\epsilon_e}
\newcommand{\epsB  }{\epsilon_B}

\newcommand{\alSSA }{\alpha_\mathrm{SSA}}
\newcommand{\SSA   }{\mathrm{SSA}}

\newcommand{\Msun       }{M_\odot}
\newcommand{\Mej        }{M_\mathrm{ej}}
\newcommand{\lumL       }{\mathcal{L}}

\newcommand{\dayunit    }{\mathrm{day}}
\newcommand{\daysunit   }{\mathrm{days}}
\newcommand{\cm         }{\mathrm{cm}}
\newcommand{\cms        }{\mathrm{cm\ s^{-1}}}
\newcommand{\kms        }{\mathrm{km\ s^{-1}}}

\newcommand{\gccunit    }{\mathrm{g\ cm^{-3}}}
\newcommand{\speclumunit}{\mathrm{erg~s^{-1}~Hz^{-1}}}
\newcommand{\GHz        }{\mathrm{GHz}}

\newcommand{\lp }{\left( }
\newcommand{\rp }{\right)}
\newcommand{\lsb}{\left[ }
\newcommand{\rsb}{\right]}

\shorttitle{SNe Ia and Shell CSM}
\shortauthors{Harris, Chelsea E., Nugent, Peter E., Kasen, Daniel N.}

\begin{document}


  \title{Against the Wind: Radio Light Curves of Type Ia Supernovae 
         Interacting with Low-Density Circumstellar Shells}

  \author{Chelsea E. Harris}
      \affil{Astronomy Department, University of California Berkeley, Berkeley, CA}
      \email{chelseaharris@berkeley.edu}
  \author{Peter E. Nugent}
      \affil{Astronomy Department, University of California Berkeley, Berkeley, CA}
      \affil{Lawrence Berkeley National Laboratory, Berkeley, CA}
  \and
  \author{Daniel N. Kasen}
      \affil{Astronomy Department, University of California Berkeley, Berkeley, CA}
      \affil{Lawrence Berkeley National Laboratory, Berkeley, CA}


\begin{abstract}
    For decades, a wide variety of observations spanning the radio through 
    optical and on to the x-ray have
    attempted to uncover signs of type Ia supernovae (SNe Ia) interacting 
    with a circumstellar medium (CSM).
    The goal of these studies is to 
    constrain the nature of the hypothesized SN Ia mass-donor companion. 
    A continuous CSM is typically assumed when interpreting observations of 
    interaction. However, while such models have been successfully
    applied to core-collapse SNe, the assumption of continuity 
    may not be accurate for SNe Ia, as shells of CSM could be formed by
    pre-supernova eruptions (novae).
    In this work, we model the interaction of SNe with a spherical, low density, 
    finite-extent CSM and create a suite of synthetic radio synchrotron 
    light curves.
    We find that CSM shells produce sharply peaked light curves, and identify
    a fiducial set of models that all obey a common evolution
    and can be used to generate radio light curves for
    interaction with an arbitrary shell.
    The relations obeyed by the fiducial models can be used to deduce
    CSM properties from radio observations; we demonstrate this
    by applying them to the 
    non-detections of SN 2011fe and SN 2014J. 
    Finally, we explore a multiple shell CSM configuration and describe
    its more complicated dynamics and resultant radio light curves.
    
\end{abstract}


\section{Introduction}
  \label{sec:intro}
     
	A type Ia supernova (SN Ia) is the explosion of a carbon-oxygen 
    white dwarf \citep{2011Natur.480..344N,2012ApJ...744L..17B}. It is hypothesized that in order to explode, these 
    stars need to gain mass from a companion star via some mass transfer 
    mechanism, be it Roche lobe overflow, winds, merger, or something more exotic. 
    The zoo of possible mass transfer mechanisms results in a similar 
    plethora of companion candidates from white dwarfs to red giants. 
    To date, there has never been a direct detection of a normal SN Ia companion \citep{2015Natur.521..328C};
    but there are indirect methods of constraining the companion identity. 
    One method is to search for signs of the ejecta impacting a circumstellar 
    medium (CSM) produced during mass transfer, pre-supernonova outbursts, or stellar winds.
      The ejecta-CSM interaction radiates across a broad range of the spectrum,
    from x-ray to radio.
    The CSM composition and structure
    will reflect the nature of the progenitor system. We refer to any supernova 
    with CSM interaction as an ``iSN''. 
    Observations of iSNe are often interpreted through the
    \citet{C82a} self-similar model of shock evolution. This model has
    successfully been used to fit radio light curves of core-collapse radio supernovae 
    \citep[e.g.][]{UsesC82a-1, UsesC82a-2, UsesC82a-3},
    supporting the hypothesis that the signal is from interaction of the SN ejecta
    with a wind. The key to the model's success in that case is that the CSM 
    is extensive and close to the stellar surface, so
    the shock evolves to the self-similar solution before observations begin.

    In the case of iSNe Ia, the density profile of the CSM is unconstrained by both
    observations and theory.
    Observations so far have been used to derive upper limits on the density
    of {\it continuous} CSM \citep{Panagia++2006, Margutti++2012, CSM12, Silver++2013}.
    However, SN Ia progenitor models come in numerous flavors, many of which do
    not predict a wind-profile CSM or even a continuous CSM.
    For instance, a dense CSM shell close to the progenitor white dwarf 
    has been considered to explain high velocity features in SN Ia spectra
    \citep{Ger++04, Silver++15}. 
    Several progenitors models suggest that nova outbursts will sweep CSM  into a shell
    of small radial extent
    \citep[e.g.][]{Patat++2011, WVS2006, MooreBild2012}. 

    It is the purpose of this work to characterize the radio light curves 
    of supernovae interacting with CSM in a shell.
    We restrict our studies to the case of 
    low-density CSM such that the interaction is in the adiabatic, optically thin limit.
    We also assume, unless stated otherwise, that there is negligible
    CSM interaction prior to interaction with the CSM shell.
    These assumptions are consistent with the fact that the majority of
    SNe Ia show no signs of interaction: if present, CSM must be of low density.
    
    The paper is organized as follows. In the next section we describe our
    model assumptions and guide the reader through our fiducial model set, 
    which is defined by the ratio  of
    the CSM and ejecta densities at the initial contact point, 
    motivated by \citet{C82a}.
    In \S~\ref{sec:shell_LCs} we present our calculated radio synchrotron
    light curves for the fiducial set, demonstrate their similarity, and
    provide a system for paramterizing these light curves such that 
    CSM properties could be easily deduced from a well-observed light curve.
    Then in \S~\ref{sec:high_vel_ej} we relax the assumption that the
    CSM and ejecta densities start in the ratio that defines the fiducial
    set, and show that the fiducial model relations can still be used 
    to approximate CSM properties even for these models.
    With that in mind, in \S~\ref{sec:obscomp} 
    we use our fiducial model light curves to analyze the probability of 
    detecting shells in recent radio observations, given that there are 
    gaps between observations.
    Finally, in \S~\ref{sec:multishell} we illustrate that a system with 
    two shells will, upon collision with the second shell, 
    produce a light curve that is significantly different from the single-shell 
    light curves.
    The appendices provide the derivations of the radiation equations 
    used in this work.


\section{Description of Models and Underlying Assumptions}
  \label{sec:mods}
    
    In this section we briefly describe the hydrodynamic code used for this study
    and describe the ``fiducial model set''.
    
    To model the shell interactions, we used the 1D spherically symmetric 
    Lagrangian hydrodynamics code of  \cite{RothKasen2015}.  
    The code solves the Euler equations using a standard
    staggered-grid Von Neumann-Richtmyer technique, and includes an artificial 
    viscosity term to damp out post-shock oscillations \cite[see e.g.][]{Castor2007}.  
    Although this code can treat radiative transport via Monte Carlo methods, 
    we did not exercise the radiation capabilities in this paper.  
    Instead our calculations are purely hydrodynamical with an adiabatic 
    index of an ideal gas,  $\gamma = 5/3$, as radiative loses are small for the 
    CSM densities considered here. 
    Figure \ref{fig:paramspace} shows that our models indeed do not trap radiation
    and do not cool, based on the equations provided 
    in Appendix \ref{app:tscales}.
    
    The exact number of resolution elements (``zones'') used in these calculations 
    varied by model but was of order $\sim 10^4$. 
    As described below, the ejecta density profile has an ``inner'' and ``outer'' region; 
    the outer ejecta interacts with the CSM and is important for the hydrodynamics.
    We asserted that the outer ejecta and CSM zones had the same mass resolution. 
    In most models we include a region of ``vacuum'' outside the CSM shell, and in some 
    models we include zones of inner ejecta. These regions act as buffers to ensure that 
    the shock front does not reach the grid boundary and create numerical artifacts. 
    The vacuum region is constant density at least a factor of $10^3$ less dense than the 
    rest of the model, and the inner ejecta is much denser than the rest; therefore 
    resolution matching is infeasible in these regions and they are assigned $3\times10^3$
    zones each when present.
    
    The hydrodynamics code outputs files containing tables of gas properties at 
    user-defined, linearly-spaced time intervals. 
    We call these output files ``snapshots'', and they represent only a subset of 
    the numerical time steps taken by the code.
    Note that, in general, snapshots are taken at different times since explosion
    for each model. Therefore we linearly interpolate if we wish to know a model
    property at a specific time, introducing a small interpolation error. 
    
    \subsection{The Fiducial Model Set}
  \begin{figure}[t]
  \centering
  \includegraphics[width=3.4in]{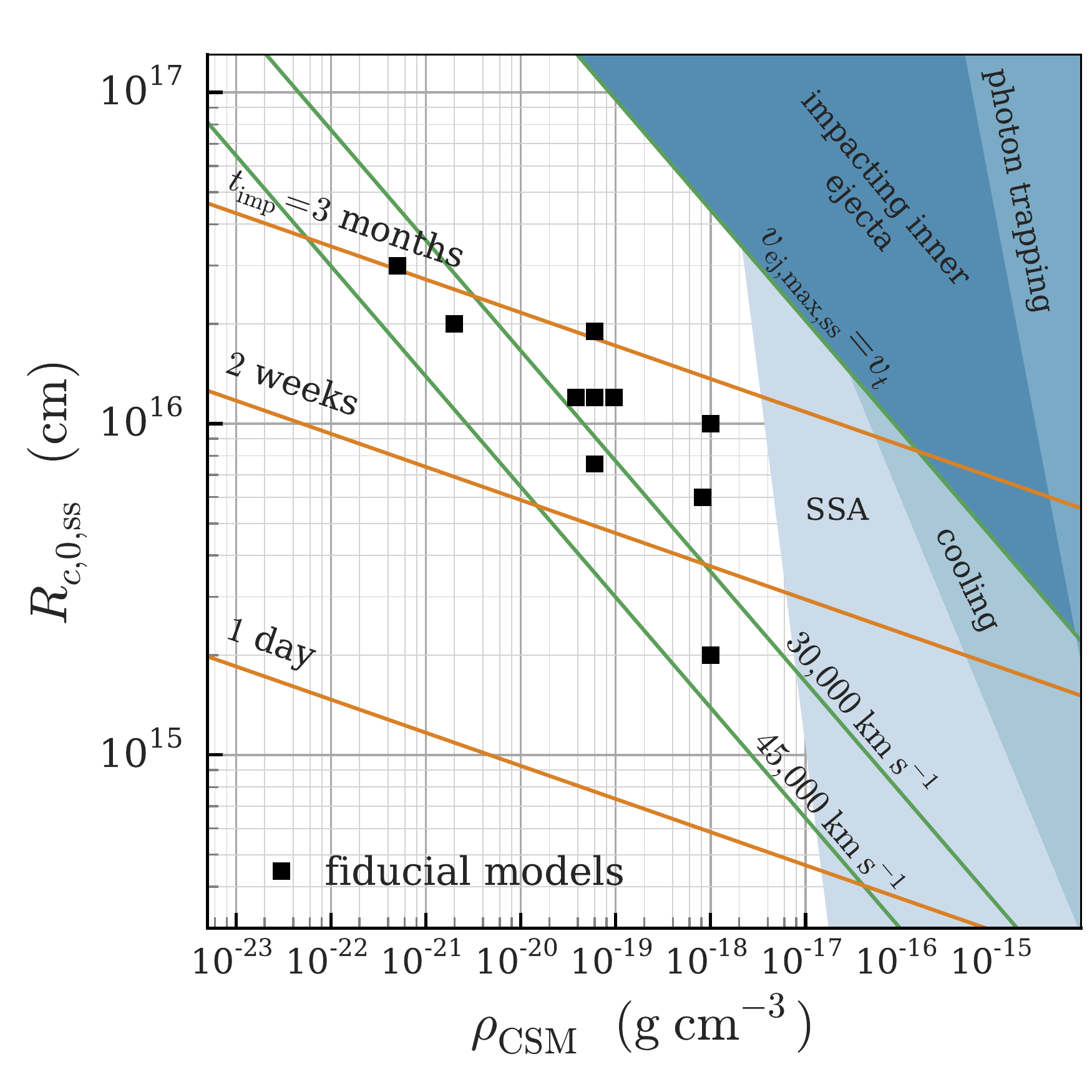}
  \caption{ A visual representation of the fiducial models presented in 
  			this work (black squares) and assumptions 
            (Appendices \ref{app:tscales} and \ref{subapp:ssa}),
            depicted in the $\rhocsm - R_{c,0,\dubs}$ plane, 
            where $\rhocsm$ is the CSM density and $R_{c,0,\dubs}$ is the initial
            contact discontinuity radius for a fiducial model (Equation \ref{eqn:Rcss}).
            Note that models of different shell fractional width (Equation \ref{eqn:f})
            occupy the same position in this space.
            Lines of constant impact time, $\timp$ (orange), and maximum ejecta velocity,
            $v_\mathrm{ej,max,ss}$ (Equation \ref{eqn:vejss}; green),
            are derived from Equation \ref{eqn:Rcss}.
            Shaded regions show where our assumptions would be violated:
            when the shock region is optically thick to electron 
            scattering and traps photons ($\tdiff \geq 0.01\tdyn$),
            the shocked gas cools ($\tcool \geq 100\tdyn$), 
            the CSM impacts dense inner ejecta instead of outer ejecta layers 
            ($v_\mathrm{ej,max,ss}>v_t$, Equation \ref{eqn:v_t}), 
            or the light curve is altered by synchrotron self-absorption 
            (SSA; $\tau_\mathrm{SSA}>1$ for $T=10^9$ K and $\nu = 4.9~\GHz$, 
            Equation \ref{eqn:tau_SSA}). 
  }
  \label{fig:paramspace}
  \end{figure}

        Previous work on CSM interaction has used the self-similar formalism 
        of \cite{C82a} (hereafter, C82). 
        In that model, freely expanding ejecta described by a 
        power law density profile $\rhoej = g^n t^{n-3} r^{-n}$ runs into stationary
        CSM with density $\rhocsm = q r^{-s}$. 
        The notations of these expressions are as in C82: $g$ and $q$ are scaling
        parameters, $t$ is the time since explosion, $r$ the distance from the supernova
        center, and $n$ and $s$ are the power law indices describing the interacting media.
        The self-similar evolution allows various quantities, including shock position 
        and velocity, to be simply calculated at any time. 
        This formalism has been applied to various core-collapse supernova data with 
        favorable results, and for this reason, as well as its ease 
        of use, the C82 framework is prevalent in studies of iSNe. 
        The well-deserved popularity of this model has 
        motivated its use as a foundation for the ``fiducial model set''
        of this study. 

        Following C82, we adopt power law density profiles
        for both the ejecta and the CSM.
        For the ejecta, we use a broken power law profile
        \citep{ChevFran1992}, which provides a reasonable approximation to 
        the structure found in hydrodynamical models of SNe.  
        For SN Ia models,  \cite{Kasen2010} suggest $\rho \propto r^{-1}$
        interior to the transition velocity ($v_t$) and $\rho \propto r^{-10}$
        exterior.
        We are primarily interested in the outer ejecta ($v>v_t$), 
        which has a density profile
        \begin{eqnarray}
            \rhoej(r) = \frac{0.124 \Mej}{(v_t t)^3} \lp \frac{r}{v_t t} \rp^{-10}, \label{eqn:rhoej} \\[8pt]
            v_t = (1.014 \times 10^{4}~ \kms)~ (E_{51}/M_\mathrm{c}),
            \label{eqn:v_t}
        \end{eqnarray}
        where $M_\mathrm{c} = \Mej/M_\mathrm{Ch}$ is ejecta mass in units of the 
        Chandrasekhar mass and $E_{51} = E/(10^{51}~ \mathrm{erg})$ is the explosion energy 
        (in this work, we take $\Mej=1.38~\Msun$ and $E_{51}=1$).  We assume that the
         power-law
        profile of the ejecta extends to a maximum velocity, $v_{\rm ej, max}$, above which
        the density drops off precipitously.  
         The ejecta is initially freely expanding with uniform temperature $T_\mathrm{ej}=10^4$ K;
       the value of the preshock ejecta temperature 
        does not play a  significant role in the evolution of the shocked gas.

        The CSM is taken to be confined to a shell with an inner radius, $R_{c,0}$,
        and width 
                $\Delta R_\CSM$. We define the {\it fractional width},  $f_R$, as 
         \begin{equation} 
            \Delta R_\CSM \equiv f_R R_{c,0} 
            \label{eqn:f} ~.
        \end{equation}
        In this work a ``thin'' shell is a shell with $f_R = 0.1$ and 
        a ``thick'' shell has $f_R = 1$.
        For example, \citet{MooreBild2012} predict $f_R \approx 0.1$ for the  shells
        swept up by nova outbursts.    
        In our calculations, we assume that there is no CSM interior to  $R_{c,0}$.

        The CSM at the start of our simulation (the time of impact) 
        is constant density ($s=0$), 
        constant velocity ($v_\CSM=1~\kms$), and isothermal ($T_\CSM=10^3$ K).
        The chosen value of $v_\CSM$ is arbitrary and 
        unimportant as long as it is much less than the ejecta velocities.
        Observations of variable narrow absorption features in SN Ia spectra
        demonstrate the possibility of moderate velocities, 
        such as $v_\CSM \sim 65~\kms$ in PTF 11kx  \citep{D12}.

        The ejecta impacts the shell at time $\timp$ after explosion.
        In our set of fiducial models, we set initial conditions such that the
        ratio of $\rhocsm$ to $\rhoej$ at the contact 
        discontinuity is a fixed value.
        Physically, our motivation is that 
        the interaction typically becomes most prominent once 
        $\rhoej \sim \rhocsm$.         
        The value we use is $\rhocsm/\rhoej = 0.33$ so that the initial contact
        discontinuity radius $R_{c,0}$ corresponds to the C82 contact discontinuity
        radius at the time of impact (Equation 3 of C82)
        \begin{eqnarray}
            R_{c,0,\dubs} &=& \lsb A_R \lp \frac{g^n t^{n-3}}{q} \rp \rsb ^{1/(n-s)} \\
                          &=& \lp 0.041 \Mej v_t^7 \rp^{0.1}~ \timp^{0.7}~ \rhocsm^{-0.1} 
                 			\label{eqn:Rcss} \\
                          &=& (5.850\times10^{14}~\cm)~
                            \lp \frac{\timp}{\dayunit} \rp^{0.7}
                            \rho_{\CSM,-18}^{-0.1},
        \end{eqnarray}
        where $A_R=0.33=\rhocsm/\rhoej$ (for $n=10, s=0$) and 
        $\rho_{\CSM,-18}=\rhocsm/(10^{-18}~\gccunit)$;
        because of this connection to the C82 self-similar solution, 
        we use the subscript ``$\dubs$'' in this work to note when this constraint
        on the initial conditions is in place.
        Although our calculations do not assume a self-similar structure, restricting our
        initial setup in this way turns out to produce
        a family of light curve models that are amenable
        to simple parameterization.        
        We explore calculations with different initial conditions in \S~\ref{sec:high_vel_ej}.
               
        Equation~\ref{eqn:Rcss} shows that, within the fiducial model set, choosing 
        $R_{c,0}$ and $\rhocsm$ fixes the impact time $\timp$, and hence
        the maximum ejecta velocity $v_{\rm ej, max} = R_{c,0}/\timp$.
        Figure \ref{fig:paramspace} illustrates the values of $R_{c,0,\dubs}$ and 
        $\rhocsm$ shown in this work, and shows that the fiducial model set obeys the 
        assumptions of transparency and lack of radiative cooling 
        (quantified in \ref{app:tscales}).    

  \begin{figure*}[t]
  \centering
  \includegraphics[width=5in]{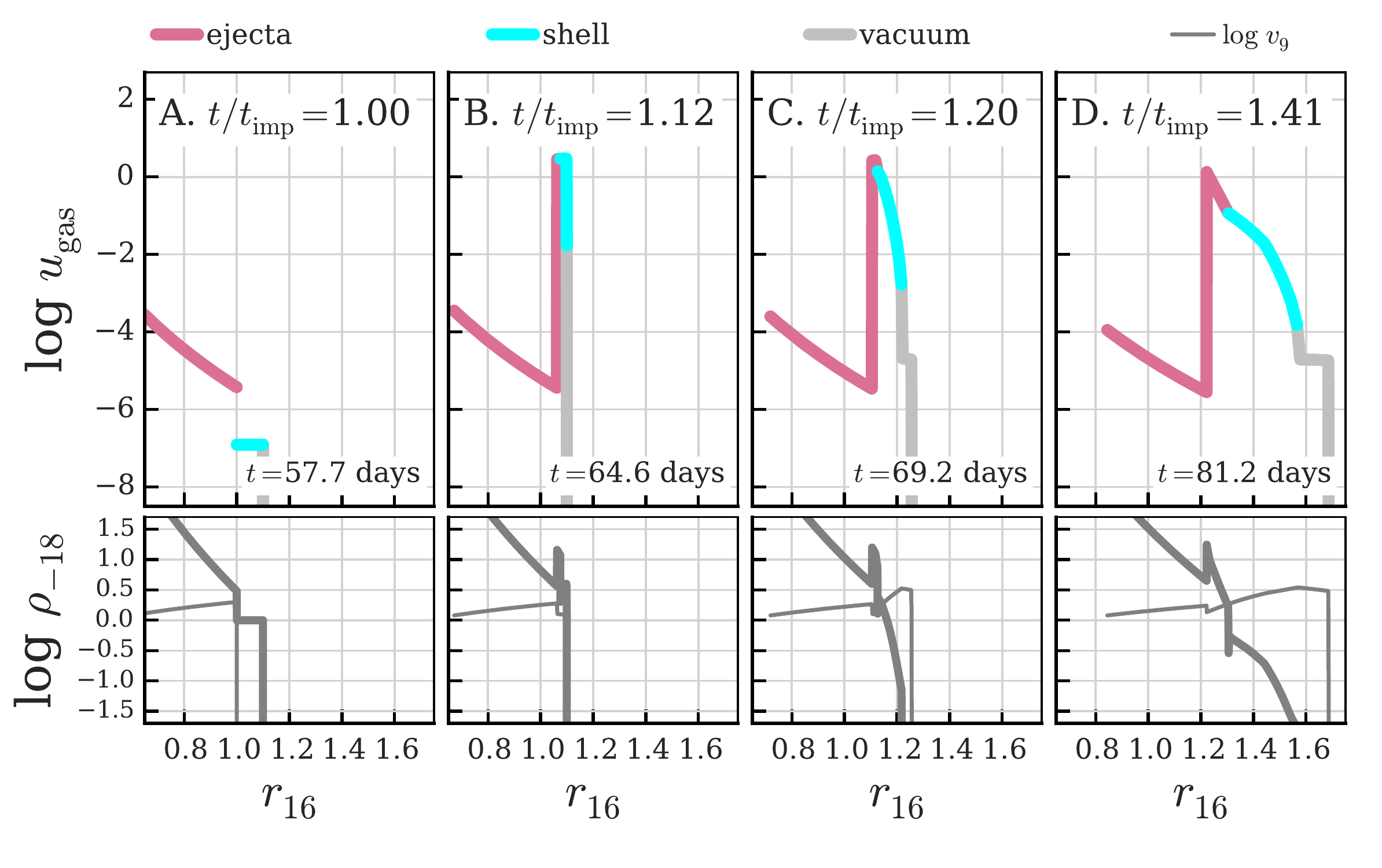}
  \caption{The evolution of internal energy density in $\mathrm{erg~cm^{-3}}$
  		  (log scale, $\log \ugas$; top), 
          mass density per $10^{-18}~\gccunit$ 
          (log scale, $\log \rho_{-18}$; bottom; thick),
          and velocity per $10^9~\rm cm~s^{-1}$ (log scale, $\log v_9$; bottom; thin)
          as a function of radius per $10^{16}~ \mathrm{cm}$ (linear scale, $r_{16}$) 
          in a fiducial thin shell impact model ($f_R=0.1, \vejmax = 20,047~\kms, 
          \rhocsm=10^{-18}~\gccunit, R_{c,0}=10^{16}~\mathrm{cm}$).
          In the top panels, different regions of the model
          are color-coded: ejecta (pink), CSM (cyan; radiating region), 
          and vacuum (grey). These regions do not mix in our simulations.
          The times shown are the initial state (A); 
          the moment of shock breakout (B); 
          after shock breakout, before all of the CSM has been accelerated (C); 
          and after significant expansion (D). 
          Labels refer to both the time relative to impact ($\timp$) and time since
          explosion ($t$).
          We see that energy density, which dominates the shell emissivity 
          (Equation \ref{eqn:physj_p3}), is high while the shock front is in the 
          shell and drops dramatically after shock breakout.
  }
  \label{fig:hydro_ev}
  \end{figure*}

        Figure \ref{fig:hydro_ev} shows the evolution of the energy and mass density profiles
        for a supernova impacting a typical $f_R=0.1$ 
        shell model.
        Immediately before impact (panel A), the CSM has low energy density, a flat density profile,
        and is moving slowly.  The interaction creates a forward shock
        moving into the CSM, and a reverse shock propagating backwards
        into the ejecta.
        The forward shock eventually reaches the outer
        edge of the shell at a time we call ``shock breakout".
        At this time (panel B), the dynamics have not yet reached a self-similar state; the internal
        energy density of the gas, $\ugas$, is high and the CSM has been accelerated to 
        nearly the ejecta speed.
        Shock breakout accelerates the CSM such that shortly afterward (panel C), 
        the mass and energy densities have dropped drastically and  the 
        outermost CSM has reached speeds over $30,000\ \kms$.  
        At a time 1.41 times the impact time (panel D), the velocity profile has approximately returned to the 
        $v\propto r$ of free expansion. 
        The mass and energy density profiles will thereafter decrease
        according to adiabatic free expansion.

        The impact radius constraint $R_{c,0}=R_{c,0,\dubs}$ (Equation \ref{eqn:Rcss}) 
        defines a fiducial set of single-shell interaction models. 
        As we will see in the next section, these models generate
        a family of light curves from which it is possible to deduce CSM properties.

\section{Radio Synchrotron Light Curves for Fiducial Model Set}
\label{sec:shell_LCs}

  \begin{figure*}[t!]
  \centering
  \includegraphics[width=6in]{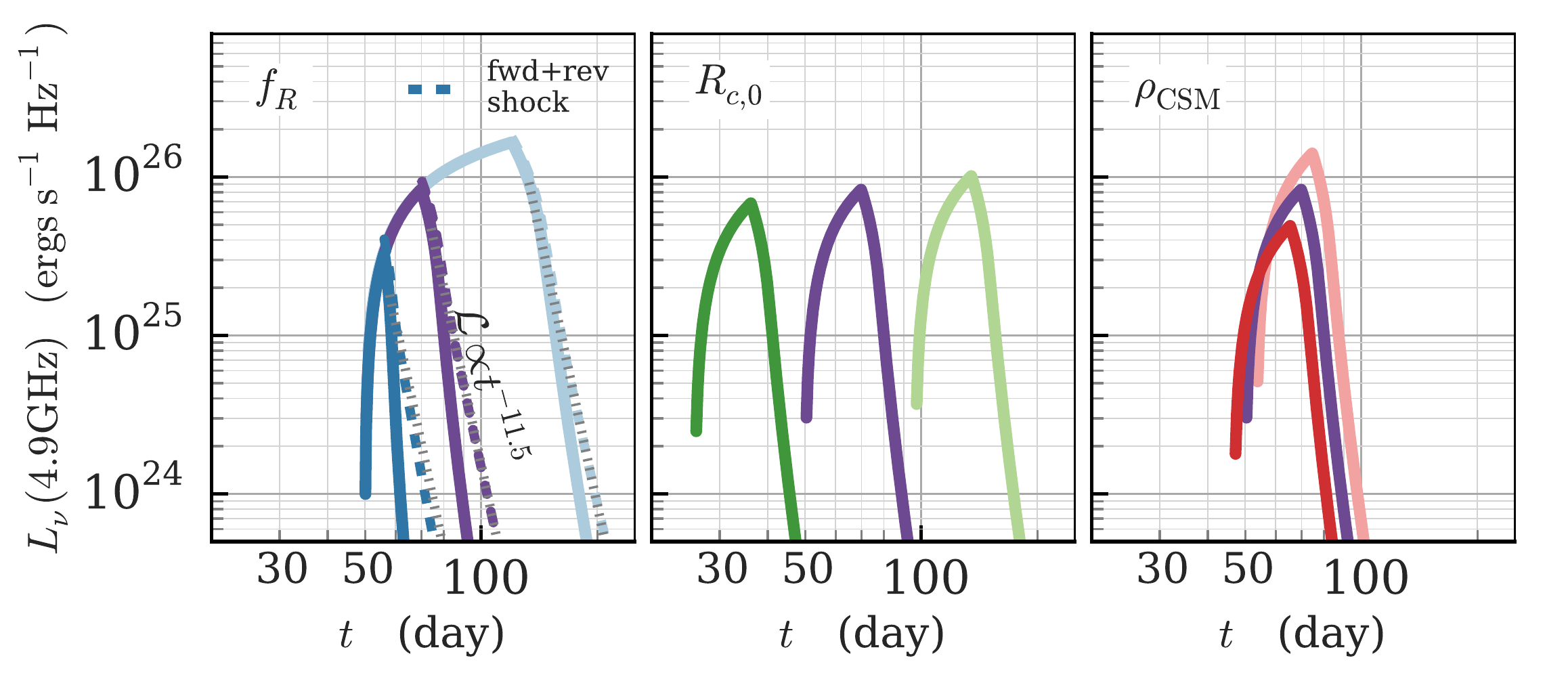}
  \caption{Dependence of 
          the fiducial model radio light curves (at 4.9 GHz) on physical parameters.
          {\it Left panel:} Effect of varying the shell fractional widths 
          from $f_R= \Delta R_{\rm csm}/R_{c,0} = 0.1,~ 0.316,~ 1.0$. 
          The light curve peak occurs at shock breakout, so wider 
          shells reach higher luminosities and have broader light curves.
          The dashed lines show light curve calculations that include  
          emission from the reverse shock region after peak. 
          The dotted grey lines 
          show the empirical fit described in \S~\ref{ssec:par_risefall}.
          {\it Center panel:} Effect of varying the initial impact radius 
          from $R_{c,0}=7.6\times10^{15},~ 1.2\times10^{16},~ 1.9\times10^{16}~\mathrm{cm}$.
          Increasing $R_{c,0}$ increases the time of impact, but causes only a
          a small increases in the peak luminosity.
		  {\it Right panel:} Effect of varying CSM density from 
          $\rhocsm=3.8\times10^{-20},~ 6\times10^{-20},~ 9.6\times10^{-20}~
          \gccunit$. Higher densities result in a higher peak luminosity, with a small 
          effect on impact time (as expected from Equation \ref{eqn:Rcss}). }
  \label{fig:phys_lc}
  \end{figure*}

    Radio emission is a tell-tale sign of interaction and commonly used 
    to study astronomical shocks.
    Here we present radio synchrotron light curves of the models described
    in the previous section, using the equations given in Appendix \ref{subapp:emissivity}.
    These models are restricted to low densities where synchrotron self-absorption is
    not important (Figure \ref{fig:paramspace}).
    Calculating light curves requires that we parameterize the fraction of postshock energy
    in magnetic fields, $\epsilon_B$, and in  non-thermal relativistic electrons, $\epsilon_e$.  
    We assume constant values typically used in the literature, $\epsilon_B = 0.1, 
    \epsilon_e = 0.1$, but note that these quantities are a main source of uncertainty in predicting
    the radio emission. 
    Calculating x-ray and optical line signatures of our models
    is an exciting possibility for future work, but here we wish to focus 
    on illustrating the  radio emission produced by shell interaction and how it differs from
    the commonly applied self-similar model.

    We assume that the CSM is of solar composition and that the
    conditions in the shocked ejecta are such that 
    only the fully-ionized shocked CSM (temperature exceeding $10^4$ K)
    contributes to the synchrotron emission \citep{ChevFran2006, Warren++2005}. 
     Representative light curves are shown in Figure \ref{fig:phys_lc}
    for interactions with shells of various widths, impact radii, and CSM densities.
    The light curves initially rise as the shock moves through the shell and increases the 
    volume of shocked material (the emission region). 
    The light curves reach a sharp peak, then rapidly decline. 
    The peak occurs at shock breakout, which we define as the moment 
    the radius of the forward shock front reaches the outer radius of the CSM.
    The rapid decline can be attributed to the plummeting internal energy of the gas
    as the shell suddenly accelerates (Figure \ref{fig:hydro_ev}, Equation \ref{eqn:jnu_synch_3p}).
    Changing only the shell width changes the time of shock breakout and therefore
    wider shells -- in which breakout necessarily occurs later -- 
    produce broader, brighter light curves.
    Thinner shells follow the rise of the thicker shells until 
    the time of shock breakout, as expected since the shock evolution should be the same while
    inside the shell.

    In the left panel of Figure \ref{fig:phys_lc} we show how including the
    contribution of the reverse shocked ejecta affects the radio light curves.  
    We use the same radiation parameters for the ejecta as for the CSM 
    (except we assume metal-rich gas so $Z/A\approx0.5$ in the calculation of electron density; see
    Appendix \ref{subapp:emissivity}). 
    We find that under these assumptions the 
    reverse shock contributes negligibly (about 10\%) to the light curve before shock breakout
    but becomes dominant at later times, when it has a much higher energy density 
    than the CSM (see panel D of Figure \ref{fig:hydro_ev}). However, we stress
    that it is unclear whether the conditions in the postshock ejecta are the same
    as in the postshock CSM; emission from the postshock ejecta could be dimmer than we have
    calculated here if, for instance, this region has a lower value of $\epsB$. 
    Because of the uncertainty in how to treat the ejecta relative to the CSM, 
    in this work we do not include the ejecta when calculating the radio emission unless
    we explicitly state otherwise.
    
    The distance of the shell from the supernova primarily affects the impact time, as one
    may expect, and has a small effect on the peak luminosity reached; the density of the 
    shell primarily affects the luminosity and has a small effect on the impact time. 
    When fractional width is held constant as is done when we vary impact radius and 
    shell density, we see that the light curves have the same shape and are simply
    shifted around in $\log t - \log \lumL$ space, as can be seen in the middle and 
    right panels of Figure \ref{fig:phys_lc}.

\begin{figure}[t]
  \centering
  \includegraphics[width=3.5in]{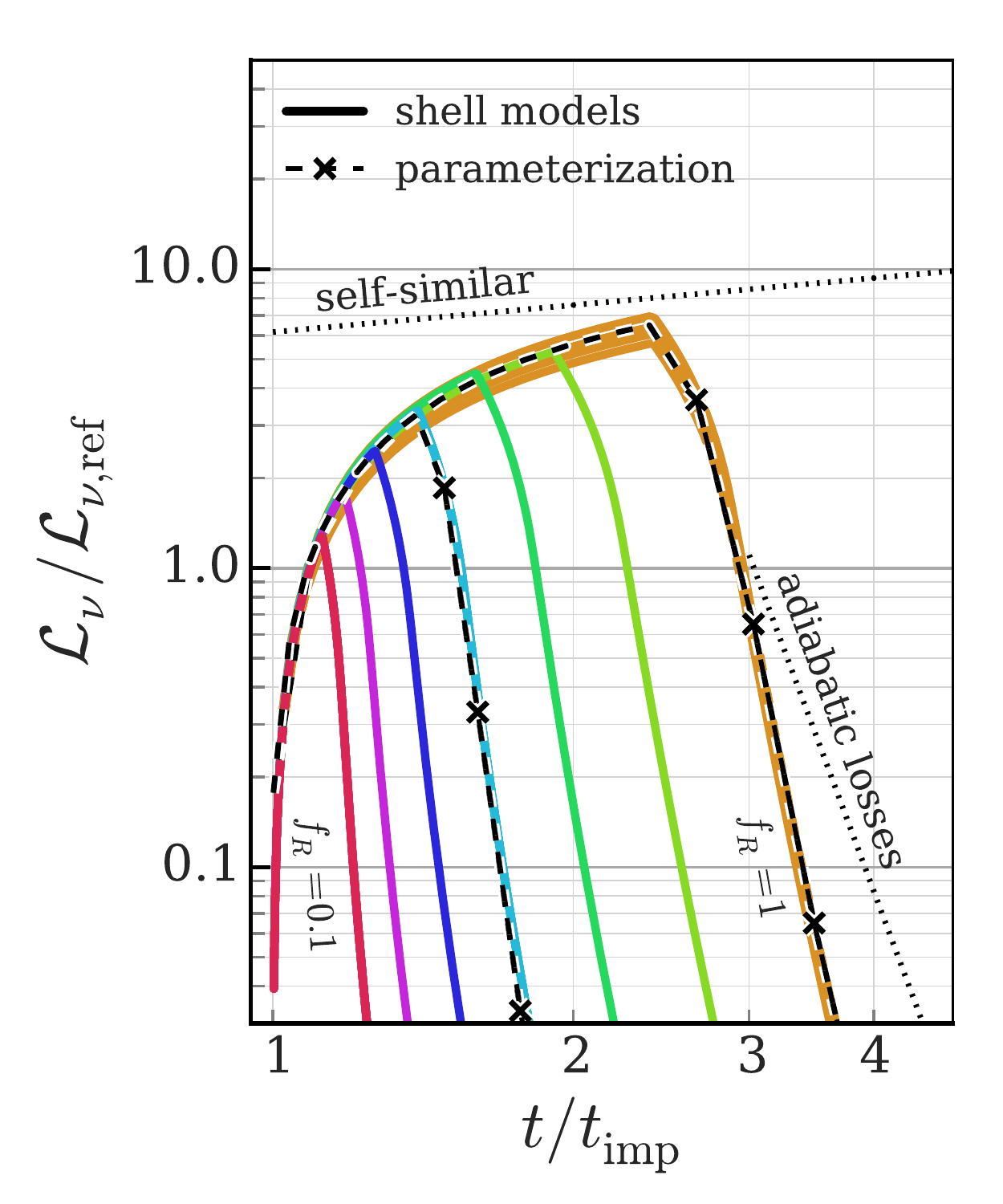}
  \caption{Light curves at 4.9 GHz with  
          time normalized to the time of impact, $\timp$, and  
          luminosity normalized to the luminosity at $1.09\timp$ 
          ($\lumL_{\nu,\mathrm{ref}}$), when all light curves are rising.
          Curves are color coded by shell fractional width, $f_R$ = $\Delta R_{\rm csm}/R_{c,0}$,
          with logarithmic spacing between $f_R=0.1$ and $f_R=1$.  
          Multiple models of varying $\rhocsm$ and $\timp$ are shown for 
          $f_R=0.1,0.316,1.0$ (red, cyan, and orange, respectively)
          to demonstrate that the normalized light curves
          are nearly identical. 
           Reconstructed light curves based on the empirical parameterization described in 
          \S \ref{sec:shell_LCs} are shown for  $f_R=0.316, .01$ 
          (black dashed lines), and 
           provide an excellent approximation to the hydrodynamical model light curves.
          The slopes predicted for self-similar evolution 
          ($\lumL \propto t^{0.3}$, Appendix \ref{subapp:selfsimLC_calc})
          and for free expansion with adiabatic energy loss 
          ($\lumL \propto t^{-9}$, Appendix \ref{subapp:adiabaticLC})
          are shown as dotted lines with arbitrary normalization; 
          they do not capture the shape of these light curves.
          }
  \label{fig:radio_lc_samp}
  \end{figure}

    In Figure \ref{fig:radio_lc_samp} we show the light curves of our fiducial 
    suite of numerical models (having different fractional widths, CSM densities, and impact times)
    normalized by their impact time and early-time luminosity.  
    This view confirms that changing the impact radius and CSM density really does just 
    shift the light curves in log-space and details of the light curve shape are governed by 
    $f_R$ alone.

    For reference, we show in Figure \ref{fig:radio_lc_samp} the slope of the 
    radio light curve predicted by the self-similar model  
    (Equation \ref{eqn:Lss_tscale}).  
    While at later times the shell light curves
    approach the shallow slope of the self-similar model, at most times this
    slope does not describe the
    light curve. We also show the decay expected after peak from a simple 
    analytic model that accounts for adiabatic losses in 
    freely expansion gas (Equation \ref{eqn:Ladiab}).  The analytic 
    curve is shallower than the decline from our models,
    as the models accelerate and expand faster than free
    expansion immediately after shock breakout.
 
    The similarity of the fiducial model set light curves is rooted in fact that the 
    models all have the same value of $A_R$,
    the initial ratio between the pre-shock CSM and ejecta density at the 
    contact discontinuity (Equation \ref{eqn:Rcss}).     
    Our next step is to quantitatively define the family of light curves 
    so that any light curve can be cheaply reconstructed for a given $f_R$ 
    and peak luminosity $\lumL_p$.

    \subsection{Parameterization of Light Curve Peak Time}
    \label{ssec:par_peak}

      First, we quantify how the fractional width of the shell, $f_R$, 
      affects the timescale for the light curve to reach peak, $\tpeak$. 
      The peak occurs when the forward shock 
      radius ($R_1$) is equal to the outer shell radius, 
      $[1+f_R] R_{c,0}$ (see Equation \ref{eqn:f}).
      In the C82 self-similar solution, $R_1 = 1.131 R_c$ and $R_c/R_{c,0} = (t/\timp)^{0.7}$
      for our choices of $n=10$ and $s=0$.  This predicts that the time of peak is related to $f_R$ through
      \begin{equation}
          f_R + 1 = 1.131 \lp \frac{\tpeak}{\timp} \rp^{0.7} 
          ~~ \Leftrightarrow ~~
          \frac{\tpeak}{\timp} = 0.839 (f_R+1)^{1.43} .
      \end{equation}    

      If instead the shock is assumed to move at constant velocity (as is true in the early evolution 
      before self-similarity is reached)  $R_1/R_{c,0} = t/\timp$ so $\tpeak/\timp = 1+f_R$.

\begin{figure}[t]
  \centering
  \includegraphics[width=3.4in]{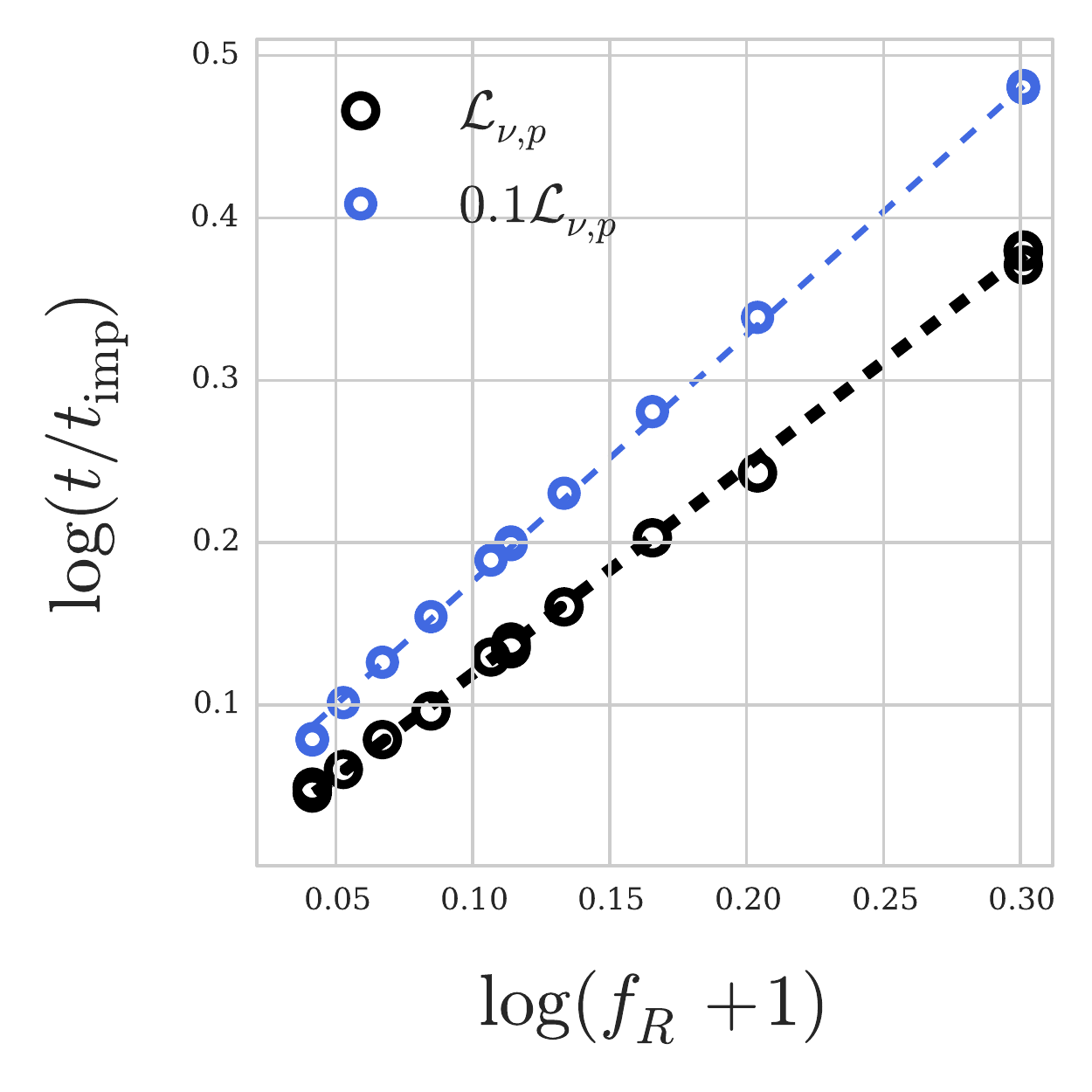}
  \caption{Time of peak (black) and time to fall to 10\% $\lumL_{\nu,p}$ (blue) 
  		  normalized to impact time 
          versus the fractional width of the shell, $f_R$ (Equation \ref{eqn:f})
          for our fiducial model suite (open circles).
          Broader shells (i.e., higher $f_R$) have broader light curves 
          (as illustrated in Figure \ref{fig:radio_lc_samp}).
          This trend is well fit by a power law (dashed lines), which 
           forms the basis for the parameterization of a 
          fiducial model light curve.  Fit parameters for the power 
          law of Equation \ref{eqn:plawfit} are given in Table \ref{tab:fits}.
          }
  \label{fig:radio_tpeak}
  \end{figure}

      Figure \ref{fig:radio_tpeak} shows $\tpeak/\timp$ versus $(f_R+1)$ for our suite of numerical calculations, 
      which we find are well fit with the power law 
      \begin{equation}
              \frac{t_p}{\timp} = 1.11 \lp \frac{(f_R+1)}{1.1} \rp ^{1.28}    ~.
              \label{eqn:plaw_peak}
      \end{equation}
      Unsurprisingly, the exponent $\alpha=1.28$ lies between the 
      self-similar ($\alpha=1.43$) and free-expansion ($\alpha=1$) predictions.

      We find empirically that the peak luminosity depends on $R_{c,0}$ and $\rhocsm$
      in the way predicted by self-similar evolution (Equation \ref{eqn:Lss}) 
      but with a different normalization than predicted by that relation.  
      For the $f_R = 1$ light curves the best fit peak luminosity is
      \begin{eqnarray}
          \left. \lumL_{\nu,p} \right|_{f_R=1} &\approx&
              (1.9\times10^{30}~ \speclumunit)~ 
              \lp \frac{\nu}{\mathrm{GHz}} \rp^{-1} \nonumber \\
              && \times \rho_{\CSM,-18}^{8/7}~ R_{c,0,16}^{3/7}
          \label{eqn:empirL-wide}
      \end{eqnarray}
      where $R_{c,0,16}=R_{c,0}/(10^{16}~\cm)$ is the scaled initial contact discontinuity radius.
      The peak luminosity of thinner shells -- which as we see in Figure \ref{fig:phys_lc} 
      peak at a lower luminosity because shock breakout occurs sooner --
      can be found with a parameterization of the light curve rise, given below. 

    \subsection{Parameterization of Light Curve Rise and Fall Times}
    \label{ssec:par_risefall}
          
      To fit the light curve rise, we note that all light curves 
      follow the same rise shape up until the time of peak. 
      A function of the form $\lumL = a - b/t$ fits the rise of $f_R=1$ 
      light curves well, with
      \begin{equation}
          \lumL_\nu(t) = 1.705~ \left. \lumL_{\nu,p} \right|_{f_R=1} 
                          \left[ 1 - 0.985 \lp \frac{t}{\timp} \rp^{-1} \right ]      
      \end{equation}
      so using Equation \ref{eqn:empirL-wide} the rise of fiducial model light curves is 
      \begin{eqnarray}
          \lumL_\nu(t) &=& (3.2\times10^{28}~ \speclumunit)~ 
              \lp \frac{\nu}{\mathrm{GHz}} \rp^{-1} \nonumber\\
              &&\times \rho_{\CSM,-18}^{8/7}~ R_{c,0,16}^{3/7}
              \left[ 1 - 0.985 \lp \frac{t}{\timp} \rp^{-1} \right ] ~.
          \label{eqn:asymL}
      \end{eqnarray}

      If desired, the peak luminosity of {\it any} fiducial model can be found by 
      substituting Equation \ref{eqn:plaw_peak} in to Equation \ref{eqn:asymL}
      \begin{eqnarray}
          \lumL_{\nu,p,\dubs} &\approx& (3.2\times10^{28}~ \speclumunit)~ 
                                        \lp \frac{\nu}{\mathrm{GHz}} \rp^{-1} \nonumber \\ 
                              &&\times  \rho_{\CSM,-18}^{8/7}~ R_{c,0,16}^{3/7}
                                        \left[ 1 - (1+f_R)^{-1.28}\right] 
          \label{eqn:empirL}                     
      \end{eqnarray}
      
      To describe the light curve decline, we choose four characteristic times along the
      fall: the time to fall a factor of $10^{-1/4} (\sim50\%),~10^{-1},~10^{-2},$ and $10^{-3}$
      of $\lumL_{\nu,p}$. 
      As demonstrated in Figure \ref{fig:radio_tpeak} for the $\lumL = 10^{-1}\lumL_{\nu,p}$, 
      case, these characteristic times follow a power law in $(f_R+1)$,
      \begin{equation}
          \frac{t}{\timp} = Q \lp \frac{f_R+1}{1.1} \rp^\alpha
          \label{eqn:plawfit}
      \end{equation}
      with fit parameters $Q$ and $\alpha$ given in Table \ref{tab:fits}. 
      Between characteristic points, we interpolate light curves linearly
      in $\log(\lumL)-\log(t)$ space (Figure \ref{fig:radio_lc_samp}). 

  \begin{deluxetable}{lcc}[ht!]
  \tablewidth{200pt}
  \tablecaption{Parameters for Equation \ref{eqn:plawfit}
  \label{tab:fits}} 
  	\tablehead{\colhead{Luminosity} & \colhead{$Q$} & \colhead{$\alpha$}}
      \startdata
      \sidehead{peak}
      $\log \lumL_{\nu,p}$        &  1.11 & 1.28        \\
      \sidehead{waning}
      $\log \lumL_{\nu,p} - 0.25$ &  1.16  & 1.39       \\
      $\log \lumL_{\nu,p} - 1$    &  1.22  & 1.52       \\
      $\log \lumL_{\nu,p} - 2$    &  1.32  & 1.62       \\
      $\log \lumL_{\nu,p} - 3$    &  1.48  & 1.70       
      \enddata
    \tablecomments{Using $Q$ and $\alpha$ in Equation \ref{eqn:plawfit}
    			   will return the $t/\timp$, as a function of $f_R$,
    			   at which a given luminosity (first column) is reached.}
  \end{deluxetable}

      Putting the parameterized rise and fall expressions together, 
      one can create an approximate light curve
      for any peak luminosity and shell width; the reconstruction will be a good 
      approximation to the full numerical calculation, as shown in Figure \ref{fig:radio_lc_samp}.      
      Finally, as mentioned previously, the  decline of the light curve is modified when
      one includes a contribution from the reverse shock (Figure \ref{fig:phys_lc}). 
      Empirically, we find that the light curve at times 
      later than the characteristic point $10^{-1/4}\lumL_{\nu,p}$ declines like 
      $\lumL_\nu \propto t^{-11.5}$, which when combined with the equation for time
      to decline to $10^{-1/4}\lumL_{\nu,p}$ (Table \ref{tab:fits}) yields
      \begin{equation}
          \lumL_{\nu,\mathrm{(f+r)}} = 0.67 \lumL_{\nu,p}~ (f_R+1)^{16}~ (t/\timp)^{-11.5}
      \end{equation}
      where the subscript ``f+r'' indicates the luminosity of the light 
      curve including both the forward and reverse shock components while 
      $\lumL_{\nu,p}$ is as given in Equation \ref{eqn:empirL}. 
      We illustrate this fit in the left panel of Figure \ref{fig:phys_lc}.

\section{Effect of Maximum Ejecta Velocity}
\label{sec:high_vel_ej}

  \begin{figure*}[t]
  \centering
  \includegraphics[width=5in]{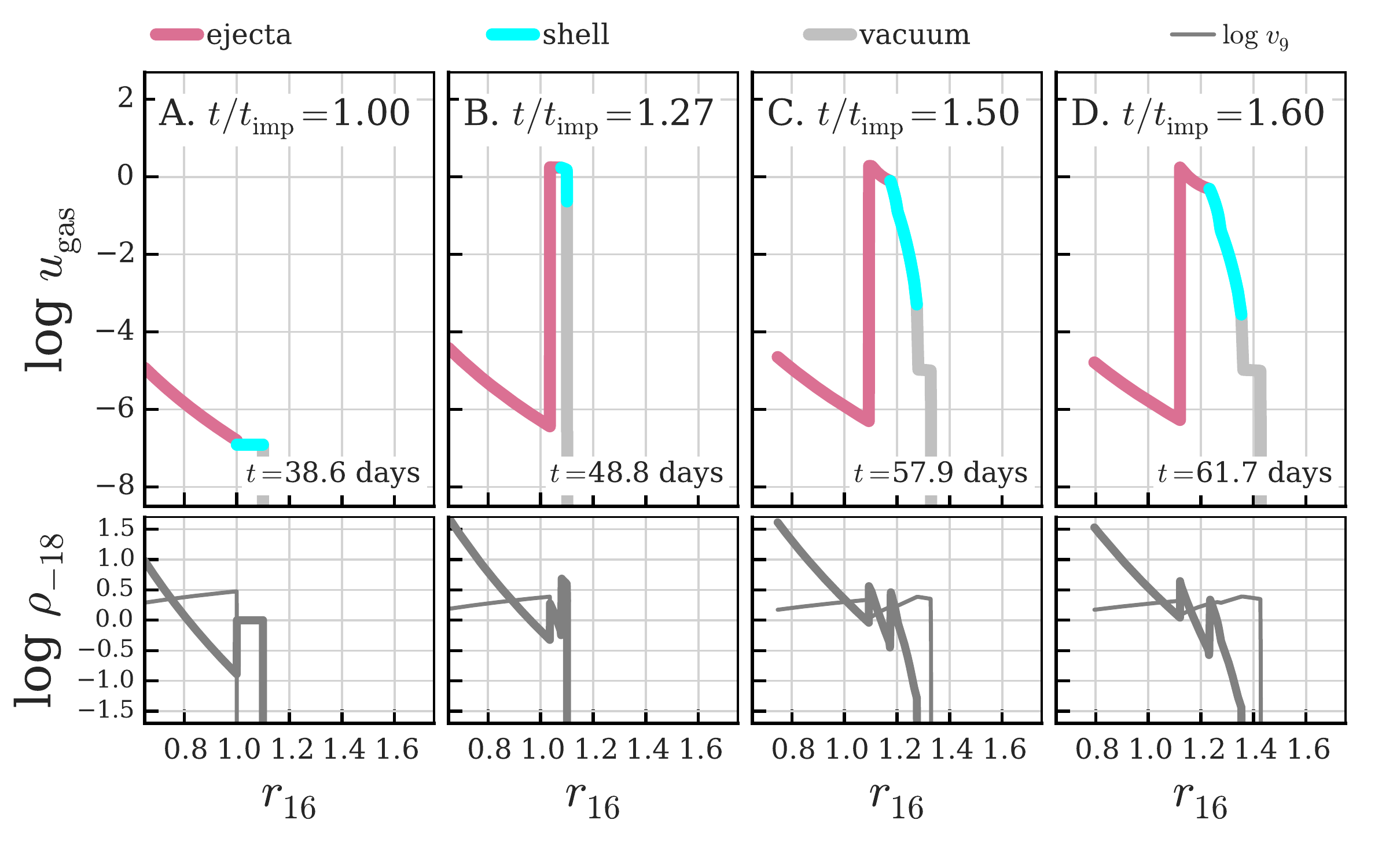}
  \caption{ Same as Figure \ref{fig:hydro_ev} but for a 
            ``high-velocity'' initial conditions model with
            $\vejmax=30,000~\kms$ ($\timp=38.6~\daysunit$) and the same 
            CSM as in Figure \ref{fig:hydro_ev}.
            Compared to a fiducial model, the ejecta is less dense
            at the initial point of contact;
            so the shock is slower, shock breakout happens later, 
            and the CSM does not gain as much energy. 
            \label{fig:hydro_highv} }
  \end{figure*}

  \begin{figure}[t]
  \centering
  \includegraphics[width=3.4in]{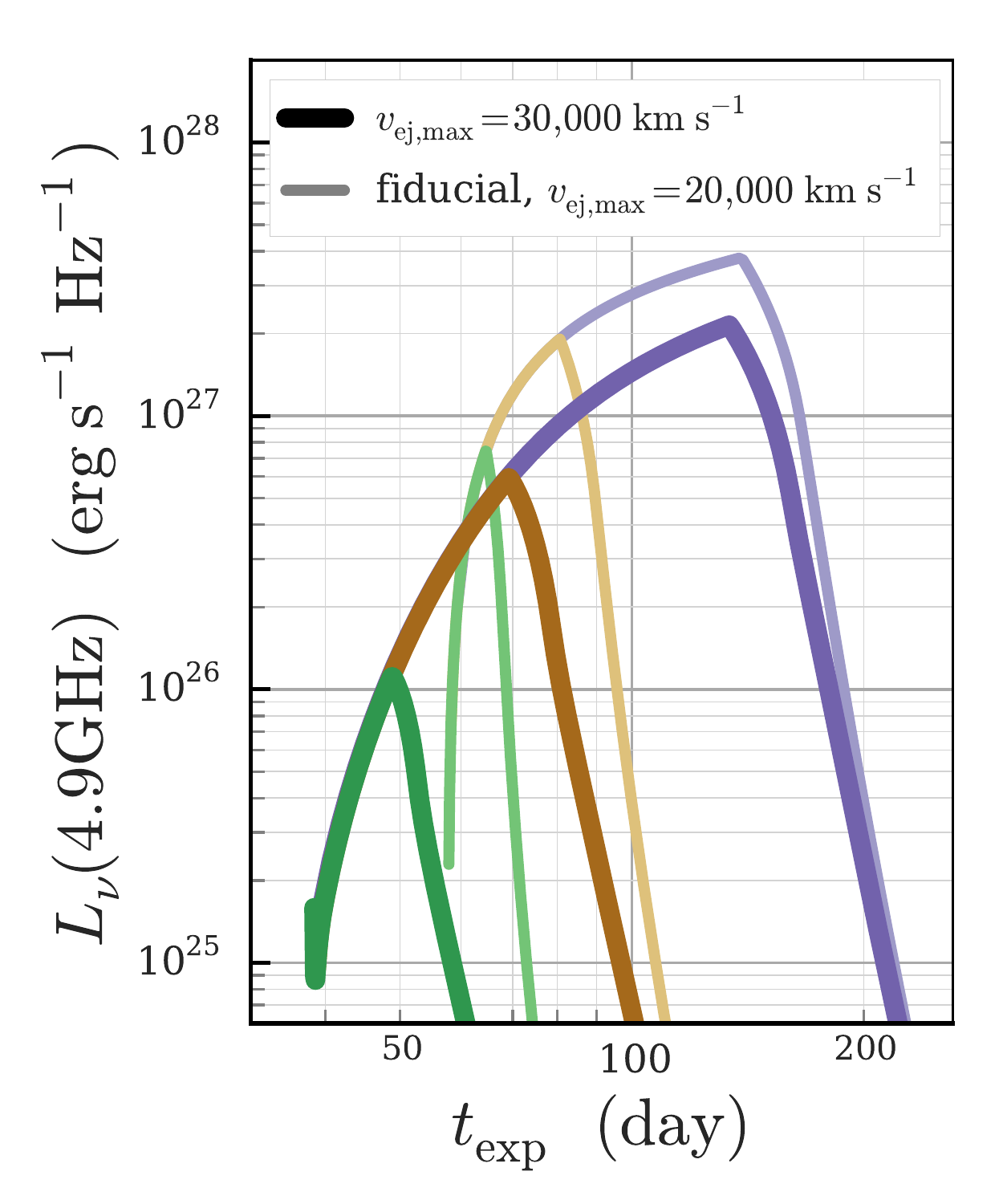}
  \caption{Light curves at 4.9 GHz, comparing fiducial
          models (\S\ref{sec:mods}; $\timp=57.7~\daysunit, \vejmax = 20,000~\kms$,
          light thin) to ``high velocity ejecta'' models (\S\ref{sec:high_vel_ej}; 
          $\vejmax=30,000~\kms$, $\timp=38.6~\daysunit$, dark thick).
          The CSM shell widths shown are:
          $f_R=0.1$ (green, narrowest light curves), $f_R=0.316$ (brown), and 
          $f_R=1$ (purple, broadest light curves).
          All models have $\rhocsm = 10^{-18}\ \gccunit$. 
          Decreasing the ejecta density relative to CSM (i.e. increasing maximum velocity) 
          lowers the peak luminosity and causes the light curve to peak at an earlier 
          time since explosion. However, the effect is negligible for thick shells, 
          and even for thinner shells the fiducial relations can estimate shell properties.
  \label{fig:radio_highv} }
  \end{figure}

  In our simulations we must choose an initial radius (equivalently, velocity) at which to 
  truncate the ejecta, $R_{c,0}$ = $\vejmax \timp$.
  The fiducial model set is defined by $R_{c,0} = R_{c,0,\dubs}$ 
  (Equation \ref{eqn:Rcss}), so the maximum velocity is 
  \begin{eqnarray}
      v_\mathrm{ej,max,ss} &=& R_{c,0,\dubs}~\timp^{-1} \\
                           &=& (2\times10^9~\cms)~ R_{c,0,16}^{-3/7}~ \rho_{\CSM,-18}^{-1/7}
      \label{eqn:vejss}
  \end{eqnarray}
  with $R_{c,0,16}=R_{c,0,\dubs}/(10^{16}~\cm)$.
  Early time spectra of normal SNe Ia show ejecta with maximum 
  velocities $\vejmax \approx 30,000~\kms$ \citep{Parrent++2012, Silver++2015}, 
  and we take $\vejmax = 45,000~\kms$ as an upper limit.
  One could constrain oneself to fiducial models in this velocity range, 
  but as illustrated in Figure \ref{fig:paramspace} that is a narrow band of 
  models.
  Considering that the ejecta velocity is one of the most important 
  constraints on SN Ia explosion models, we wanted to explore 
  the effect of ejecta with $v>v_\mathrm{ej,max,ss}$ on our light curves.
  We call these ``high-velocity models'', although they could also be thought of 
  as ``super-$A_R$'' models, as they have a higher initial CSM to ejecta density than the fiducial
  model value $A_R = 0.33$.
  Note that these models are not
  plotted on Figure \ref{fig:paramspace}, to avoid confusion regarding the reference
  lines that apply only to fiducial models. 

  Figure \ref{fig:hydro_highv} shows the hydrodynamic evolution 
  for a high-velocity model analogue 
  of the model shown in Figure \ref{fig:hydro_ev}. 
  The fiducial model of Figure \ref{fig:hydro_ev} was initialized with 
  $\rhocsm = 10^{-18}~\gccunit$ and $R_{c,0} = 10^{16}~\cm$, which implies 
  $\vejmax \approx 20,000~\kms$ and $\timp = 57.7$ days.
  The high-velocity model Figure of \ref{fig:hydro_highv} has the same CSM properties
  but $\vejmax=30,000~\kms$, thus it has $\timp \approx 38.6$ days ($A_R=1.00$).
  The high-velocity model has a lower initial ram pressure, 
  since at a fixed radius $\rhoej \propto t^{7} \propto v^{-7}$ 
  (Equation \ref{eqn:rhoej})
  and $p_\mathrm{ram}\propto \rhoej v^2 \propto v^{-5}$; 
  therefore the forward shock is slower 
  and the light curve takes a few hours longer to reach peak 
  (time of shock breakout; Figure~\ref{fig:hydro_highv}, panel B). 

  Example light curves of high-velocity models with varying $f_R$ are presented in 
  Figure \ref{fig:radio_highv}, along with their fiducial model counterparts.
  Including the higher velocity ejecta leads to light curves that are broader,
  dimmer, and peak earlier.
  As discussed, the effect on breadth is because the shock is slower so it takes 
  longer to cross the shell. 
  The dimming is also due to the lower shock speed 
  (Equations \ref{eqn:u_rhovsq} and \ref{eqn:jnu_synch_3p}). 
  That the light curves peak earlier relative to explosion time is simply 
  due to the earlier impact time.

  The offsets in peak luminosity and time of peak are more pronounced in 
  thinner shell models,
  both in absolute and relative terms. Therefore we decided to apply the
  fiducial model relations of \S~\ref{sec:shell_LCs} to see how well they
  recover shell properties from the $f_R=0.1$ high-velocity model.
  If the $\vejmax=30,000~\kms$ light curve had been observed, 
  one would calculate that $f_R=0.2$ (Equation \ref{eqn:plaw_peak}), 
  $\rhocsm = 1.2\times10^{-18}~ \gccunit$, and 
  $R_{c,0} = 7.5\times10^{15}~\cm$ (Equations \ref{eqn:Rcss} and \ref{eqn:empirL})
  -- all correct to within a factor of a few.

  Overall, we conclude that one can still apply the fiducial model relations
  to an observed light curve to reasonably estimate
  CSM properties even if the derived parameters yield a $v_\mathrm{ej,max,ss}$ 
  that is below the observed ejecta velocity, $\vejobs$. 
  Therefore, in the next section, we will take the liberty of applying the 
  fiducial relations to the analysis of iSNe Ia radio non-detections.

\section{Observational Constraints}
\label{sec:obscomp}
    In this section we use the radio observations of SN~2011fe from
    \citet[hereafter CSM12]{CSM12} and of SN 2014J from \citet[hereafter PT14]{PT++2014}
    to illustrate how our models can be used to interpret radio 
    nondetections in SNe~Ia. As detailed below, 
    we do this by calculating a suite of parameterized light curves 
    relevant to the observations, determining if each would be observed, 
    then converting these observation counts into observation
    likelihoods.

\begin{figure}[t]
  \centering
  \includegraphics[width=3.4in]{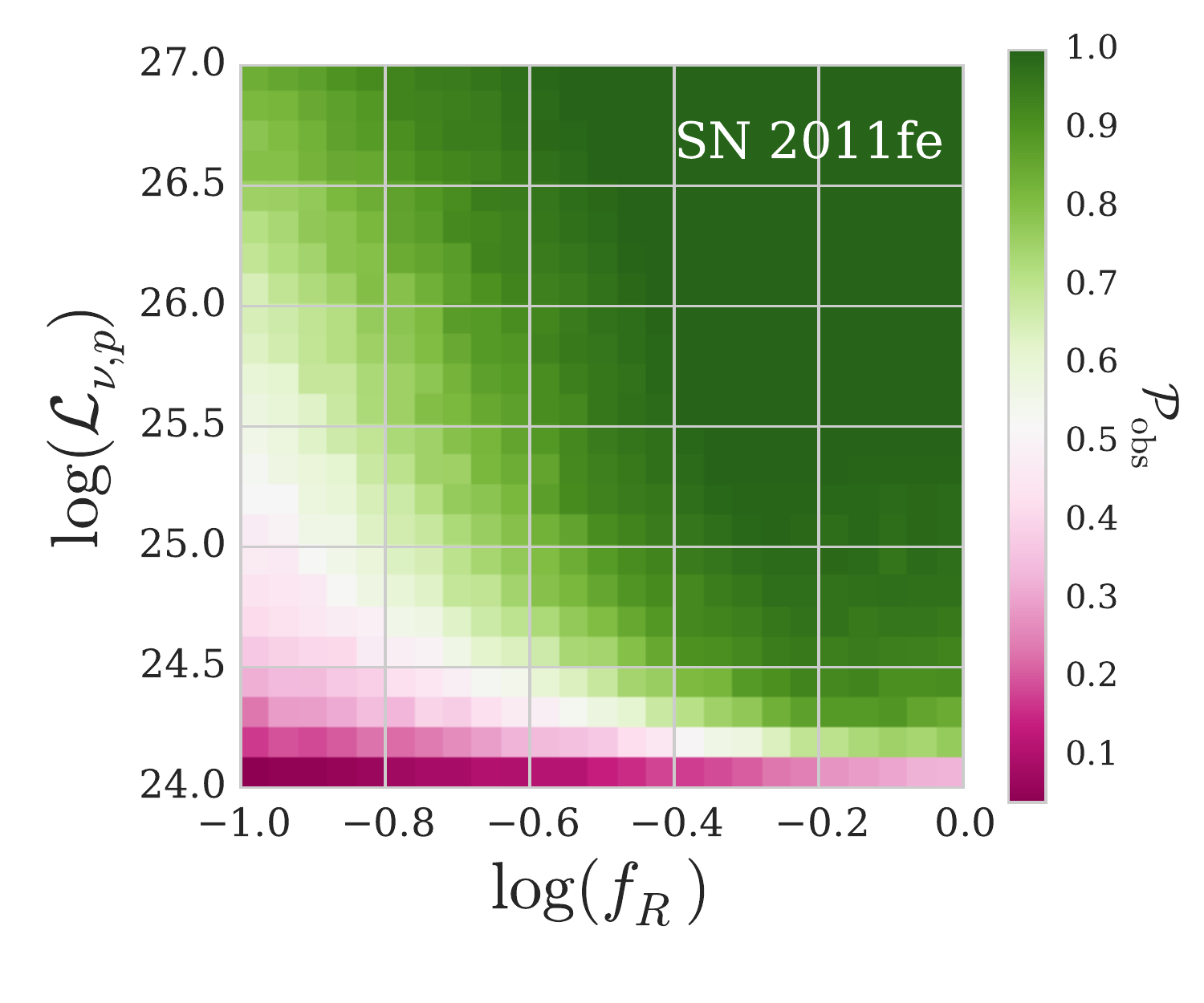}
  \caption{Probability that the CSM12 cadence would observe a 
          light curve ($\mathcal{P}_\mathrm{obs}$) 
          with peak luminosity $\lumL_{\nu,p}(5.9~\GHz)$
          from a shell of fractional width $f_R$ (Equation \ref{eqn:f}) 
          for a set of light curves that have log-uniform randomly assigned
          values of $\lumL_{\nu,p}\in [10^{24}, 10^{27})~\speclumunit,
          f_R\in [0.1,1),$ and $\timp\in[2,19)~\daysunit$.
          For a given $f_R$ and initial shell radius, 
          peak luminosity can be converted into CSM density using 
          Equation \ref{eqn:empirL} as is done in Figure \ref{fig:P_obsthin} 
          for SN 2014J.}
  \label{fig:P_obs}
  \end{figure}

    The analysis begins by defining the set of observations. 
    CSM12 observed SN~2011fe in two bands with effective frequency 
    $\nu=5.9$ GHz at times after explosion $\texp = 2.1 - 19.2~ \daysunit$, 
    with luminosity limits $\sim 10^{24}~ \mathrm{erg~ s^{-1}~ Hz^{-1}}$ 
    for each observation. 
    To interpret these observational constraints,
    we generate $10^6$ light curves corresponding to different 
    interaction models by log-uniform randomly distributed sampling of peak luminosity,
    shell width, and impact time with values in the range 
    $\lumL_\mathrm{p}(5.9~\GHz)\in [10^{24}, 10^{27})~\speclumunit$, $f_R\in [0.1,1)$,
    and $\timp \in [2,19)~\daysunit$. 
    For each light curve, we then determine if it was observed, i.e. if it was 
    above the CSM12 luminosity limit for any single observation. 
    We restrict ourselves this way since if the shell had been observed it 
    would have affected the observation schedule; but using our models one 
    could interpret stacked data sets as well.
    Finally, we bin the light curves in $f_R$ and $\lumL_p$ and in each bin calculate 
    the probability that it would have been observed 
    (the ratio of the number of light curves observed to the number sampled). 
    The probability of the light curve appearing in any single observation 
    is shown in Figure \ref{fig:P_obs}. 

  \begin{figure}[t]
  \centering
  \includegraphics[width=3.4in]{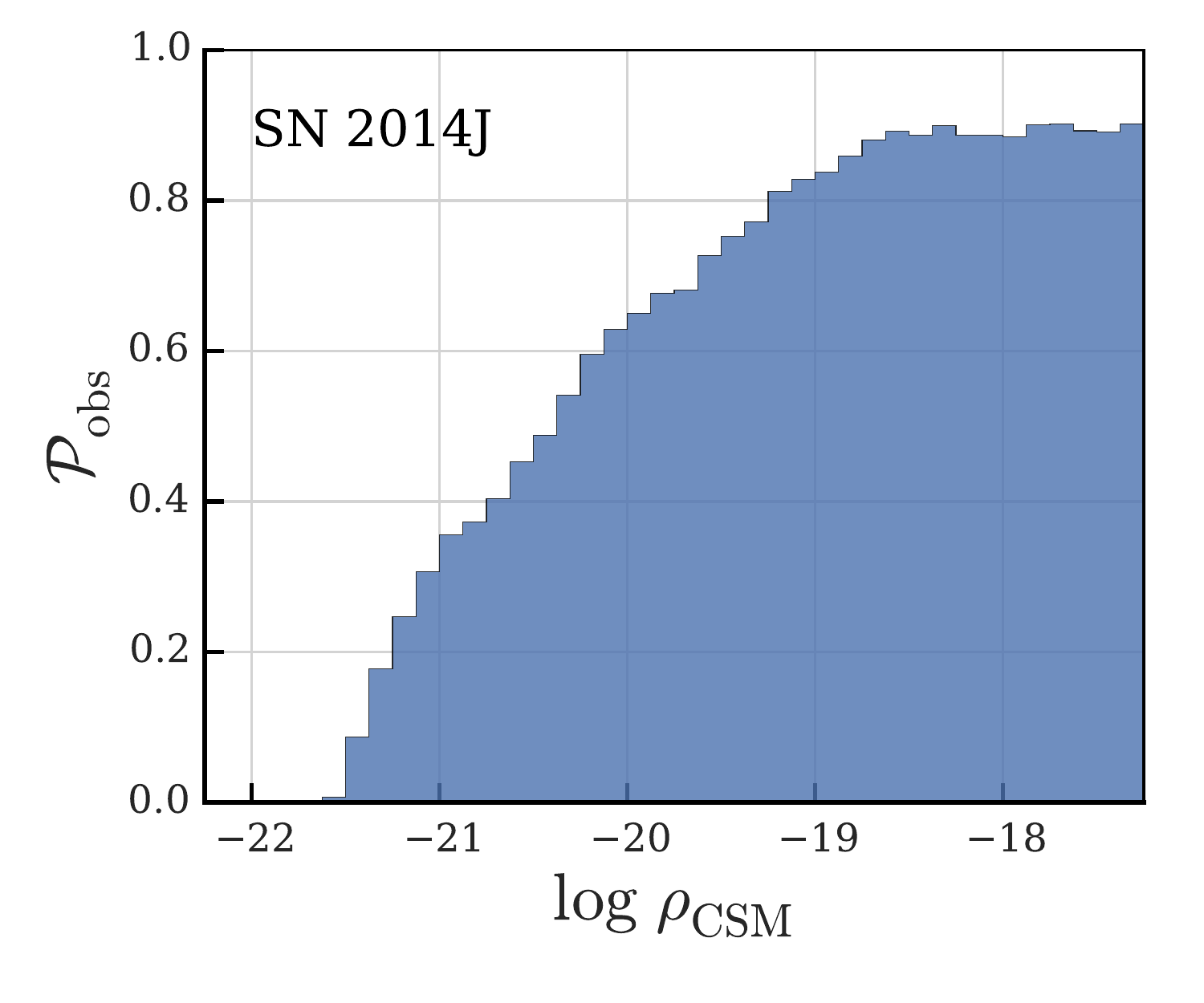}
  \caption{Probability ($\mathcal{P}_\mathrm{obs}$) that any single radio 
          observation 
          of \citet{PT++2014} would detect a thin CSM shell ($f_R=0.1$) 
          of a given mass density ($\rhocsm$) around SN 2014J 
          if impact occurred during their observation window, 8.2-35.0 days.
  		}
  \label{fig:P_obsthin}
  \end{figure}    
    
    CSM12 attempted to estimate, without using explicit shell light curve models, the probability of 
    detecting a specific CSM shell. The shell they considered had fractional width $f_R\sim0.1$,
    mass $6\times10^{-8}~\Msun$ 
    (from a wind of $\dot{M} = 2\times 10^{-7}~ \Msun~\mathrm{yr^{-1}}$),
    and sat at $R\approx 5\times 10^{15}$ cm; the authors estimated a 30\% 
    chance of detecting the shell. 
    The preshocked CSM density is 
    \begin{equation}
        \rhocsm = (4.7\times10^{-24}~ \gccunit)~ 
             M_{-8}~ R_{16}^{-3} \left[(1+f_R)^3 - 1 \right]^{-1}
    \end{equation}
    where $M_{-8} = M_{\CSM}/(10^{-8}~ \Msun)$ and $R_{16}=R/(10^{16}~\mathrm{cm})$. 
    We estimate that the shell described by CSM12, 
    which has $M_{-8}=6,~R_{16}=0.5$, and $\rhocsm \approx 7\times 10^{-22}~\gccunit$ 
    would have a peak luminosity (Equation \ref{eqn:empirL})
    $\lumL_{\nu,p} (5.9~\mathrm{GHz}) \approx 6 \times 10^{22}~\speclumunit$, 
    and would be undetected in their observations. 
    For reference we note that at our assumed distance and width, a shell would need to have
    $M_{-8} \approx 600$ to have $\sim 50\%$ chance of detection. 
    
    For SN 2014J, we will demonstrate how one can instead analyze the 
    probability of observing a shell of a given density, rather than peak luminosity.
    The observations of SN 2014J by \citet{PT++2014} span the range 8.2-35.0 days. 
    Using the observation times, frequencies, and limits from 
    PT14 Table 1, we perform the same random sampling test as for 
    SN 2011fe described above, but with a peak luminosity range 
    $10^{23-28}~\speclumunit$ at 1.55 GHz (their lowest frequency band) 
    and impact times between 2 and 35 days. 
    Isolating the $f_R=0.1$ bin, we can convert luminosity to density and 
    analyze the probability that the authors would have observed an impact
    with a shell of a given $\rhocsm$ that occurred during their observations,
    illustrated in Figure \ref{fig:P_obsthin}.
    The same shell considered for CSM12 but at $R=10^{16}~\mathrm{cm}$, 
    to account for the later observation time,
    would again escape detection at $\rhocsm \approx 10^{-22}~\gccunit$.

    As we have shown here, by using the parameterized light curves and fiducial model
    relations one can easily quantify the probability of detecting shells spanning a 
    range of $f_R$ and $\rhocsm$ for a given set of observations.
    Alternatively, one could constrain for instance the ejecta velocity 
    and $f_R$, and use Equation \ref{eqn:vejss} to analyze the likelihood of 
    detecting shells of varying radii and densities.

\section{Multiple Shell Collisions}
\label{sec:multishell}

    We have demonstrated that the light curves of ejecta impacting a single
    shell form a family of solutions that can be straightforwardly applied to 
    the interpretation of radio observations. In these simulations, we assume that
    there is negligible CSM interaction prior to the interaction being considered - 
    i.e. the space between the supernova progenitor and the CSM shell is empty. 
    However, it is natural to wonder how it would look if one shell collision 
    was followed by another: could we simply add the light curves, preserving their 
    convenient properties? In this section we perform this experiment and show that 
    the light curves from multiple shell collisions differ dramatically from
    a simple addition of two single-shell light curves. 

    Imagine a system with two thin ($f_R=0.1$) CSM shells: 
    Shell 1 at $9.2\times10^{14}$ cm with $\rhocsm=8.8\times10^{-15}~\gccunit$
    and Shell 2 at $7.7\times10^{15}$ cm with $\rhocsm=4.7\times10^{-18}~\gccunit$.
    The ejecta impacts Shell 1 at 5 days in a self-similar model, and if Shell 1 
    were not present the ejecta would impact Shell 2 at 50 days;
    therefore, models of the {\it single-shell} impact with these shells are 
    called ``1sh5'' (Shell 1) and ``1sh50'' (Shell 2). 
    The double-shell model (``2sh27'') is the same as 1sh5 until Shell 1 impacts 
    Shell 2 on day 27. 
    Note that the earlier impact time (27 days versus 50 days) is due to the fact that  
    this second interaction starts not when the {\it ejecta} impacts Shell 2 but rather when {\it Shell 1} does,
    which happens earlier since Shell 1 is exterior to the ejecta.
    The initial condition of 2sh27 is 
    the day-27 state of 1sh5 for $r<R_{1-2}$ plus the initial state of 1sh50 for 
    $r>R_{1-2}$, where $R_{1-2}$ is the contact discontinuity radius between the shells.
    The number of zones in the 1sh50 CSM was adjusted to match the mass resolution 
    of 1sh5.

\begin{figure}[t]
  \centering
  \includegraphics[width=3.4in]{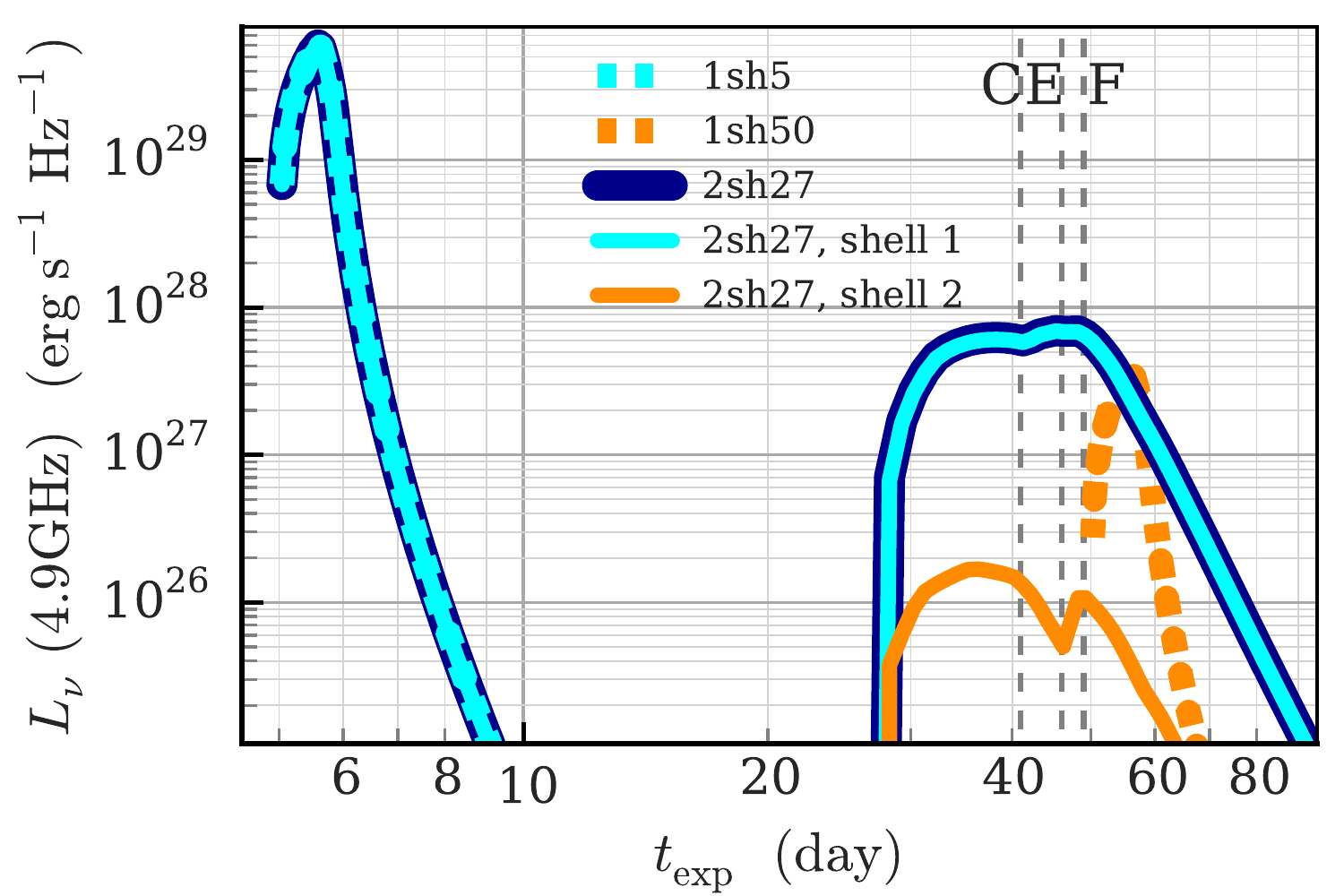}
  \caption{4.9 GHz light curve of the two-shell model 2sh27 (navy).
          It can be broken down into the radiation from Shell 1 (cyan thin) and
          Shell 2 (orange thin); after impact with Shell 2, the signal from Shell 1
          dominates in this band because the Shell 2 synchrotron spectrum peaks
          at a much lower frequency than Shell 1.
          Thick dashed lines show single-shell models:
          the light curves that would result from assuming that 
          Shell 1 (1sh5; cyan; assuming optically thin to SSA) 
          or Shell 2 (1sh50; orange) was the only shell in the system; 
          simply summing these light curves is not at all a good approximation to 
          the behavior of the two-shell system after impact with Shell 2.
          Compared to 1sh50, 2sh27 has a longer-lived and brighter light curve, with
          a plateau that last a couple of weeks and a more gradual decline. 
          Dashed vertical lines indicate important features in the light curve:
          the elbow in Shell 1 (C), elbow in Shell 2 (E), 
          and time of final descent (F), with letters corresponding to panels 
          in Figure \ref{fig:2sh_toon}.}
  \label{fig:2sh_synchLC}
  \end{figure}

\begin{figure*}[t]
  \centering
  \includegraphics[width=4.5in]{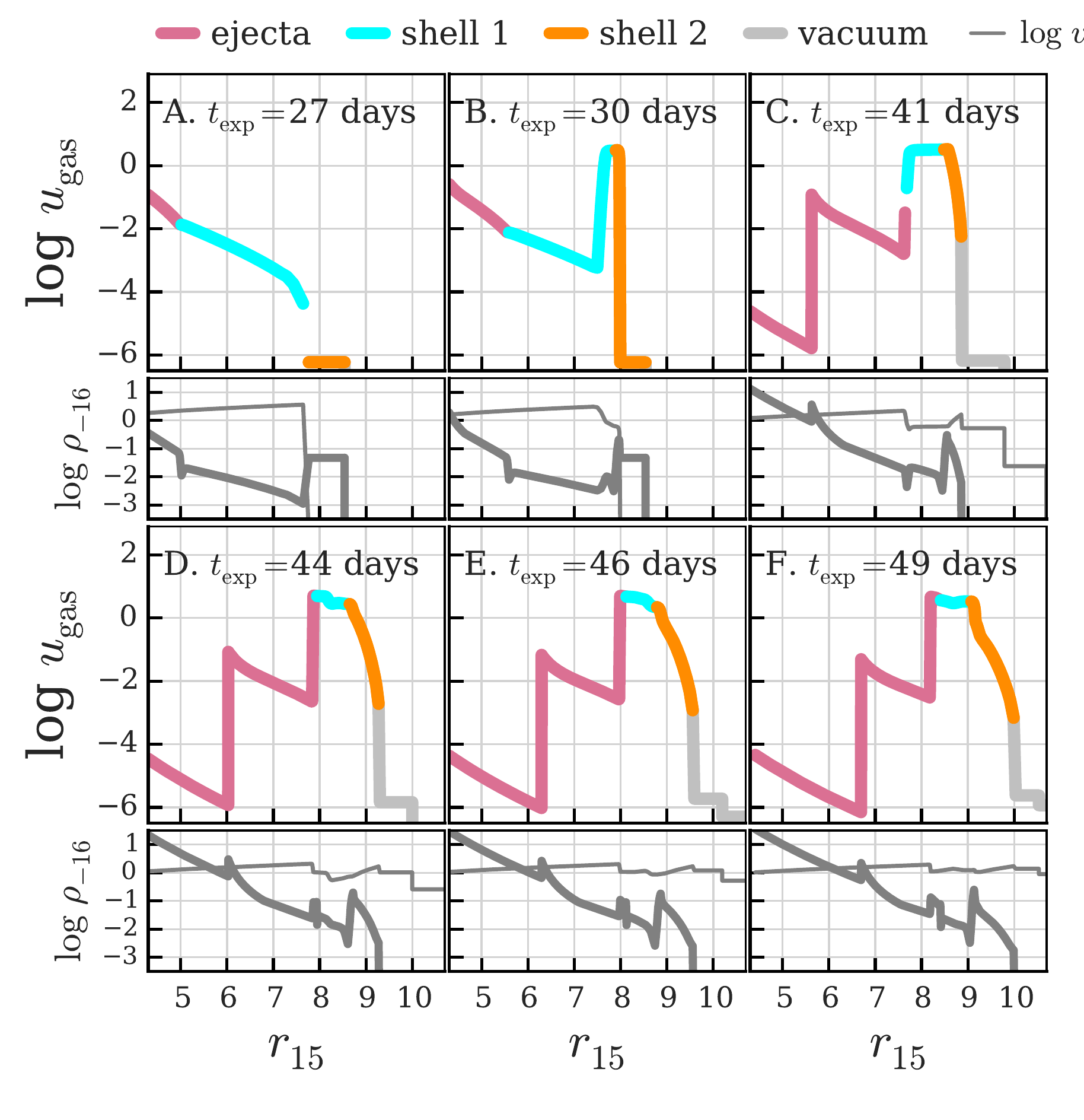}
  \caption{Like Figure \ref{fig:hydro_ev}. 
  			Snapshots of 2sh27 (double shell model) at important times in its evolution, 
            showing $\ugas$ in the ejecta (pink), Shell 1 (cyan), Shell 2 (orange), 
            and the vacuum (light grey). The energy density in Shell 1 stays high
            because of ``shock sloshing'': the recurring formation of shocks in 
            Shell 1 caused by a shock front reaching either the ejecta on the left
            or Shell 2 on the right and forming a new shock front in Shell 1 that 
            travels the opposite direction.
  }
  \label{fig:2sh_toon}
  \end{figure*}

	Figure \ref{fig:2sh_synchLC} shows that the 2sh27 light curve 
    is dramatically different than those of 1sh5 and 1sh50. 
    First, the 2sh27 light curve is strikingly flat and broad, maintaining 
    a luminosity similar to the peak luminosity of 1sh50 for nearly twenty days.
    Second, the light curve has an elbow at $\texp = 41$~days (time C; corresponding to 
    panel C of Figure \ref{fig:2sh_toon}). 
    Note that in this figure, the 1sh5 light curve would in fact be subject to
    synchrotron self-absorption, an effect that we neglect here because it is 
    not the interesting part of the evolution. 

    Since our 1-D simulations do not have mixing, we can calculate the 
    luminosity contribution from the Shell 1 and Shell 2 zones separately 
    (which zones belong to each shell is known by construction, as 
    illustrated in Figure \ref{fig:2sh_toon}).
    When separated into the contribution from each shell, we see that the light curve is 
    dominated by the signal from Shell 1. This is simply because of the chosen 
    frequency of $\nu_\mathrm{obs}=4.9$ GHz; in our model the ratio of 
    cyclotron frequency in Shell 2 to Shell 1 is
    $\nu_{c,\mathrm{sh2}}/\nu_{c,\mathrm{sh1}}\sim 2\%$,
    thus $L_\mathrm{\nu,sh2}/L_\mathrm{\nu,sh1}\sim 2\%$.
    The Shell 2 light curve also has an elbow (time E; $\texp= 44$~days)
    that occurs later than the Shell 1 elbow.
    Finally, both the Shell 1 and Shell 2 light curves start their final decline at 
    $\texp=49$~days (time F). 
    
    This complicated light curve is caused by ``sloshing'' of shocks in Shell 1
    which re-energize the gas for $\sim 27$~days, 
    the dynamical time (Appendix \ref{app:tscales})
    This ``sloshing'' can be seen in snapshots of the simulation at critical times 
    (Figure \ref{fig:2sh_toon}). 
    Remember that Shell 1 dominates the signal in the following narrative of the interaction:    
    (A) {\it Impact.} Shell 1 is much less dense than either the ejecta or Shell 2,
    and it is approaching Shell 2 at high speeds. 
    (B-C) {\it Shock moves leftward through Shell 1.} When Shell 1 hits the ``wall'' 
    of Shell 2, a shock forms. It energizes Shell 2 in a way reminiscent of single-shell
    evolution, breaking out quickly.
    (D) {\it Shock forms at ejecta-Shell 1 boundary and re-energizes Shell 1.} 
    Because of the new shock, $\ugas$ increases and the light curve has another small rise.
    (E) {\it Shock forms at the Shell 1-Shell 2 boundary, re-energizing both shells.} 
    We see that when the shock moving through Shell 1 (seen in panel D) hits Shell 2, 
    a new shock forms. The Shell 2 light curve begins a dramatic second rise and Shell 1 is able
    to maintain its luminosity. 
    (F) {\it After a dynamical time the system has had time to adjust and the light 
    curve begins to decline.}
    This panel shows a snapshot just before the decline of both light curves.
    Minor shocks continue to slosh inside shell 1 between the ejecta
    and shell 2, causing the more gradual light curve fall compared to single-shell
    models.

    This is just one possible configuration of a double-shell system and its evolution
    probably does not generalize to the whole class of such collisions;
    however, it illustrates that the evolution of these systems is distinct from
    single-shell model evolution and so produces a radically different light curve.
    Note that higher-dimensional simulations may be required for investigating
    this class further because the significance of the ``shock slosh'' 
    could be diminished by turbulence.

\section{Conclusions}
  \label{sec:conc}
    In order to quantitatively interpret the radio light curves and spectra of iSNe, 
    one needs to understand the evolution of the shocked gas. 
    One popular model for shock evolution is the 
    \citet[``C82'']{C82a} one-dimensional self-similar interaction model, which assumes
    a continuous CSM. Here we explored moving beyond these models to 
    the case of CSM shells (e.g. due to novae) in the low-density 
    case where the shocked CSM does not cool and is optically thin.
    This removes the self-similar model assumption that the interaction has
    had time to reach self-similarity (i.e. that the CSM is close to the
    progenitor star or is very extended). 
    
    We created a suite of models to explore the parameter space of shell 
    properties, focusing on shell density, thickness, and time of
    impact.
    In our ``fiducial'' models, we enforce an initial CSM-to-ejecta density ratio
    of $A_R=0.33$ at the contact discontinuity 
    (motivated by C82; Equation \ref{eqn:Rcss}) .
    If the shell were infinite, these models would evolve to the 
    C82 self-similar solution (\S~\ref{sec:mods}); however, because of 
    the finite extent of the shells, self-similarity is never reached
    in our models.

    We calculated synchrotron radio light curves for these models, 
    since this emission traces the evolution of the shocked region
    only. 
      We found a similar behavior among the radio light curves of fiducial models 
    -- they  peak at a predictable time (Equation \ref{eqn:plaw_peak}), 
    have the same shape (Equation \ref{eqn:asymL}), and 
    peak at a luminosity that can be calculated from shell characteristics
    (Equation \ref{eqn:empirL}).
    Therefore these light curves can be parameterized in a way that allows
    CSM properties to be inferred readily from an observed light curve
    (\S~\ref{sec:shell_LCs}). 
    In fact we find that shell properties can be approximated even if 
    there is a higher initial density ratio 
    (i.e. faster, lower density ejecta colliding with the shell)
    than is assumed in the fiducial set (\S~\ref{sec:high_vel_ej}).

    We then showed how one can use the fiducial models to better interpret
    radio non-detections of SNe Ia 
    by applying our work to the radio observations of SN 2011fe 
    and SN 2014J  (\S~\ref{sec:obscomp}).
    
    We presented a two-shell model to illustrate that our single shell
    light curves do not apply in that case, as a double shell light
    curve is much longer lived because of an effect that we call ``shock sloshing''
    that occurs in the first shell (\S~\ref{sec:multishell}). 
    Shock sloshing produces elbows in the light curve that may be an observable 
    signature of this behavior; exploring the diversity in multiple shell 
    collisions is an enticing direction for future work.

    Our fiducial model set is a new tool for studying the interaction of 
    SNe Ia with a CSM, with the goal of constraining
    the supernova progenitor system. With the fiducial set, one can easily 
    make observational predictions and analyze radio observations. 
    Future work will focus on calculations of the radiation transport  
    in the optical and x-ray, to show how data at those wavelengths can reinforce 
    radio constraints on supernovae interaction.
    
\section*{Acknowledgments}
    The authors thank the anonymous referee for improving
    the clarity of this manuscript. C.E.H. is supported by the 
    Department of Energy Computational Science Graduate Fellowship.

\appendix
\section{Appendix}
\label{appendix}

    \subsection{Diffusion and Cooling Timescale Estimates}
    \label{app:tscales}
        We simulate the interaction between ejecta and CSM under the 
        assumptions that (1) the gas is optically thin to electron scattering, 
        so photons can free-stream through it and 
        (2) the gas does not radiatively cool.
        To investigate the legitimacy of
        these assumptions, we can compare the timescale for dynamical
        changes ($\tdyn$) to the diffusion time ($\tdiff$) and 
        the cooling time ($\tcool$) in the forward shock region.

        The dynamical time is 
        \begin{equation}
            \tdyn = R/v \approx R_{c,0} / v_\mathrm{ej} \approx \timp 
            \label{eqn:tdyn}
        \end{equation}
        since the shocked gas is quickly accelerated to nearly the ejecta 
        velocity and the ejecta is freely expanding. 
        
        The assumption that photons can free-stream requires that the diffusion
        time be much less than the dynamical time.
        The diffusion time for electron scattering through a shocked region of width
        $\Delta R_s$ and optical depth $\tau$ is
        \begin{equation}
            \tdiff = \frac{\Delta R_s}{c} \min(\tau,1)
        \end{equation}
        Assuming that all species are fully ionized within the shocked region,
        the electron number density ($n_e$) and ion number density ($n_I$) are
        related to the mass density ($\rho$), atomic number ($Z$), atomic mass ($A$), 
        and proton mass ($m_p$) through
        $n_e = Zn_I$ and 
        $n_I = \rho/(A m_p)$, so the optical depth to electron scattering is
        \begin{equation}
            \tau = \sigma_T n_e \Delta R 
                 \approx (4\times10^{-3}) \lp \frac{Z}{A} \rp \rho_{s,-18} \Delta R_{s,16}
            \label{eqn:tau_es}
        \end{equation}        
        with $\rho_{s,-18}$ the density of the gas in the shocked region per 
        $10^{-18}~\gccunit$ and $\Delta R_{s,16}$ the width of the shocked region
        per $10^{16}~\mathrm{cm}$.
        For the estimates here, we assume the gas is hydrogen dominated so 
        $Z \approx A\approx 1$. 
        Assuming a typical value for the shock width 
        $\Delta R_s \approx 0.1 R_c$ and that
        the postshock density is related to the initial CSM density as
        $\rho_{s} \approx 4 \rho_{\CSM}$,
        \begin{equation}
           \tdiff \approx (1~ \mathrm{min})~ R_{c,16}^2~ \rho_\mathrm{CSM,-18} ~.
           \label{eqn:tdiff}
        \end{equation}
        
       The assumption that the shocked gas does not cool requires that the cooling
       time be much longer than the dynamical time.
        The cooling time for any radiative process is given by
        \begin{equation}
            \tcool = \ugas/(4 \pi j) = 1.5\ n k_B T / (4\pi j)
        \end{equation}
        where $\ugas$ is the gas internal energy density (which we have 
        assumed is that of a monatomic ideal gas) and $j$ is the 
        angle-averaged, frequency-integrated emissivity of a radiative process.
        For free-free emission of thermal electrons, the angle-integrated 
        emissivity (corresponding to $4\pi j$ in the above equation) is
        \begin{equation}
            \jff = 1.43\times10^{-27}\ Z^2 n_I\ n_e\ T^{1/2}\ \bar{g}_B
        \end{equation}
        in cgs units \citep{RL1979}. Choosing a value of 1.2 for the gaunt factor
        $\bar{g}_B$ gives an accuracy of about 20\%. Using $n_e, n_I$ as above and again
        assuming the shocked material has a density $\rho \approx 4 \rhocsm$ and 
        $A=Z=1$,
        \begin{equation}
            \tcoolff \approx (1600\ \mathrm{yr})\ \rho_{\CSM,-18}^{-1}\ T_9^{1/2}
            \label{eqn:tcool}
        \end{equation}
        where $T_9 = T/(10^9~\mathrm{K})$.
        The strong shock jump conditions for an ideal gas link the temperature and
        shock velocity as
        \begin{eqnarray}
            \frac{3}{2} \frac{\rhocsm}{\mu m_p} k T_s &=& \ugas = \frac{P}{\gamma_\mathrm{ad} - 1} 
                                     = \frac{3}{2} \lp \frac{2}{\gamma_\mathrm{ad}+1}~ \rhocsm v_s^2 \rp
                                     = 1.125 \rhocsm v_s^2 \\[10pt]
            \label{eqn:u_rhovsq}
            \; \Rightarrow \; T_9 &\approx& 9 \mu v_{s,9}^2
            \label{eqn:Tfromv}
        \end{eqnarray}
        where here $\gamma_\mathrm{ad}$=5/3 is the adiabatic index and
        $v_{s,9} = v_s/(10^9~ \mathrm{cm~ s^{-1}})$ is the shock velocity 
        per $10,000~ \mathrm{km~ s^{-1}}$. 
        For our models we expect that $v_s$ 
        is similar to the ejecta velocity so $T_9 \sim 1-10$ and thus the 
        cooling timescale is much longer than the dynamical timescale.

        For the fiducial model set, we can use Equations \ref{eqn:Rcss} and \ref{eqn:Tfromv} 
        to recast the diffusion and cooling timescales in terms of physical
        parameters. 
        Figure \ref{fig:paramspace} illustrates these constraints and 
        shows that our models obey them.

    \subsection{Luminosity Calculations}
    \label{app:synch}

    \subsubsection{Synchrotron Emissivity}
    \label{subapp:emissivity}
    
    To calculate radio synchrotron light curves, we assume a fraction $\fNT$ of electrons are accelerated into a non-thermal power law
    distribution 
    $dn_e = C \gamma^{-p} d\gamma$ (with $\gamma$ the electron Lorentz factor) with a total number density 
    $\neNT = \fNT n_e$.
    Data for  stripped-envelope core-collapse supernovae favors $p\approx3$ \citep{ChevFran2006}. 
    The normalization $C$ is given by
    \begin{eqnarray}
        \neNT &=& \int_{\gammin}^{\infty} C \gamma^{-p} d\gamma \\
        \Rightarrow\; C &=& \neNT (p-1) \gammin^{p-1} 
        \label{eqn:C}
    \end{eqnarray}
    If a fraction $\epse$ of the shock energy went into accelerating the electrons 
    (energy density $u_e$), then 
    \begin{equation}
        u_e \equiv \epse \ugas = \int_{\gammin}^{\infty} (\gamma m_e c^2)\ C\gamma^{-p}d\gamma
    \end{equation}
    which we can combine with Equation \ref{eqn:C} to show that 
    \begin{equation}
        \gammin = \frac{u_e}{\neNT m_e c^2} \lp \frac{p-2}{p-1} \rp
        \label{eqn:gammin}
    \end{equation}
    For the special case $p=3$ which we assume in this work,
    \begin{eqnarray}
        \gammin &=& \frac{1}{2} \frac{u_e}{\neNT m_e c^2}, \\[5pt]
        C &=& \frac{1}{2} \neNT \lp \frac{u_e}{\neNT m_e c^2}\rp ^2 
        \label{eqn:C_3p}
    \end{eqnarray}
    
    Assuming isotropic radiation, we can derive the synchrotron emissivity 
    of these electrons as the power density per frequency per steradian
    \begin{equation}
        j_\nu(\nu>\nu_c) = \frac{1}{4\pi} \frac{1}{d\nu} n_e(\gamma) P_e(\gamma)
        		= \frac{1}{4\pi} \frac{1}{d\nu} (C\gamma^{-p}d\gamma) P_\mathrm{syn}
        \label{eqn:j_syn_orig}
    \end{equation}
    where $P_\mathrm{syn}$ is an electron's power output in synchrotron radiation 
    \begin{equation}
        P_\mathrm{syn}(\gamma) = \frac{4}{3}\sigma_T c u_B \gamma^2 \beta^2 
        \label{eqn:Psyn}
    \end{equation}
    with $u_B$ the magnetic field energy density ($u_B = B^2/8\pi$) and $\beta = v/c$. 
    We parameterize the magnetic field energy density as $u_B = \epsB \ugas$, 
    as is common practice.
    The frequency of synchrotron radiation is related to the cyclotron 
    frequency $\nu_c = eB/(2\pi m_e c)$ as
    \begin{equation}
        \nu = \gamma^2 \nu_c
    \end{equation}
    thus
    \begin{equation}
        \frac{d\gamma}{d\nu} = (2 \gamma \nu_c)^{-1}
        \label{eqn:dgdn}
    \end{equation}
    
    Using Equations \ref{eqn:Psyn}--\ref{eqn:dgdn} in Equation \ref{eqn:j_syn_orig},
    \begin{equation}
        j_\nu(\nu>\nu_c) = \frac{\sigma_T c}{6\pi}\ C\ u_B\ \nu_c^{-1}\ 
                      \lp \frac{\nu}{\nu_c} \rp^{(1-p)/2}
          \label{eqn:jnu_synch}
    \end{equation}
    where here we have assumed that $v \sim c$, with $c$ the speed of light; 
    to relax this assumption one would multiply the above by $(1-\gamma^{-2})$.
    This is a slightly simpler version of the anisotropic equation given in 
    \citet{RL1979}, and the two are equal to within a factor of order unity
    if one takes the pitch angle terms in the latter equation 
    to be their angle-averaged value $\left<\sin\theta\right>_\theta = 2/\pi$. 
    For $p=3$, we can use Equation \ref{eqn:C_3p} and
     our parameterizations $u_B = \epsB \ugas$, $u_e = \epse \ugas$
    to write this as
    \begin{equation}
        j_\nu(\nu>\nu_c ~|~ p=3) = \frac{\sigma_T}{12\pi m_e^2 c^3}\ 
                                   \epse^2~ \epsB~ n_e^{-1}~ \ugas^3~ \nu^{-1} ~.
          \label{eqn:jnu_synch_3p}
    \end{equation}
    Assuming fully ionized hydrogen ($n_e = (Z/A)\rho/m_p$ with $Z/A=1$ as described
    for Equation \ref{eqn:tau_es}), this gives
    \begin{eqnarray}
        j_\nu(\nu>\nu_c ~|~ p=3) &=&
        (1.3\times10^{-15}~\mathrm{erg~s^{-1}~cm^{-3}~Hz^{-1}~sr^{-1}})~
        \lp \frac{\epse}{0.1} \rp^2~ 
        \lp \frac{\epsB}{0.1} \rp~\nonumber\\[8pt]
        &&\times
        \lp \frac{\rho}{10^{-18}~\gccunit} \rp^{-1}
        \lp \frac{\ugas}{10^2~\mathrm{erg~cm^{-3}}} \rp^3~ 
        \lp \frac{\nu}{\mathrm{GHz}} \rp^{-1}
        \label{eqn:physj_p3}
    \end{eqnarray}

    For our simulations, we assume the CSM is solar composition
    and that only regions with $T>10^4$ K
    radiate (and that these regions are fully ionized; recall that for times of interest the 
    radiating regions have $T\sim10^9$ K). 
    The radiating region extends from the ejecta-CSM contact radius to the forward shock radius,
    which is identified by an algorithm that finds the sharp break in the $\ugas$
    profile (see, e.g. Figure \ref{fig:hydro_ev}). 
    We calculate the luminosity by calculating the emissivity in each radiating zone 
    (zone defined in \S~\ref{sec:mods}) scaled by its peak value (at $\nu_c$), then multiplying 
    by the volume of the zone and $4\pi$ steradians,
    as is appropriate in the optically thin limit.
    For a quantitative discussion of our optically thin assumption as it applies
    to synchrotron self-absorption, see \S~\ref{subapp:ssa}.
    
    As discussed in the main text, for a few models we consider the emission from the reverse 
    shock region.  This is calculated in the same manner except that the composition 
    is assumed to dominated by species heavier than hydrogen, in which case $A/Z = 2$.

    \subsubsection{Self-Similar Evolution Light Curve}
    \label{subapp:selfsimLC_calc}

    Next we use the synchrotron emissivity to calculate the luminosity predicted from the C82
    self-similar evolution.
    In the case of constant-density CSM, we can use our model parameters (see \S~\ref{sec:mods},
    particularly Equation \ref{eqn:rhoej}) in Equation 3 of C82 to derive that 
    a self-similar shock has speed
    \begin{eqnarray}
        R_1 &=& 1.131 R_c \\
        \Rightarrow\; v_s &=& dR_1/dt = 0.792 R_c~ t^{-1} 
                          = 0.502~(\Mej~v_t^7)^{1/7}~ R_c^{-3/7} \rhocsm^{-1/7}
    \end{eqnarray}
    where $R_1$ is the radius of the forward shock front, $v_s$ is the shock velocity, 
    and as usual $\rhocsm$ is the preshock CSM density and $R_c$ is the radius of the contact
    discontinuity. 
    Substituting this into Equation \ref{eqn:u_rhovsq},
    \begin{equation}
         \ugas = 1.125~ \rhocsm v_s^2 = 0.284~(\Mej~v_t^7)^{2/7}~ R_c^{-6/7}~ \rhocsm^{5/7}~.
    \end{equation}
    Thus the emissivity (Equation \ref{eqn:jnu_synch_3p}, with $\rho=4\rhocsm$),
    assuming fully ionized hydrogen gas, evolves like
    \begin{equation}
        j_\nu = 0.092~ \frac{\sigma_T m_p}{12\pi m_e^2 c^3}\ 
               (\Mej~v_t^7)^{6/7}~ \epse^2~ \epsB~ \nu^{-1}~ 
               R_c^{-18/7}~ \rhocsm^{8/7}~ 
    \end{equation}
    and the emitting volume increases like
    \begin{equation}
        V = \frac{4\pi}{3} (R_1^3 - R_c^3) = 1.871 R_c^3
    \end{equation}
    which together produce
    \begin{equation}
        \lumL_{\nu,\dubs} \sim j_\nu V 
                          \sim 0.17~ \frac{\sigma_T m_p}{12\pi m_e^2 c^3}~
                               (\Mej~v_t^7)^{6/7}~ \epse^2~ \epsB~ \nu^{-1}~
                               R_c^{3/7}~ \rhocsm^{8/7} ~.
    \end{equation}
    For our model parameters of $\Mej=1.38~\Msun$, $v_t=1.022\times10^9~\cms$, 
    $\epse=\epsB=0.1$,
    \begin{equation}
        \lumL_{\nu,\dubs} \sim (2.3\times10^{29}~\speclumunit)~
                               \lp\frac{\nu}{\mathrm{GHz}}\rp^{-1}~
                               R_{c,16}^{3/7}~ \rho_{\CSM,-18}^{8/7} 
        \label{eqn:Lss} \\
    \end{equation}
    where $R_{c,16}=R_c/(10^{16}~\cm)$ is the current contact discontinuity radius 
    and $\rho_{\CSM,-18}=\rhocsm/(10^{-18}~\gccunit)$. From Equation \ref{eqn:Rcss}, 
    we see that $R_c^{3/7} \propto t^{3/10}$ 
    so the scaling of luminosity with time is
    \begin{equation}
        \lumL_{\nu,\dubs} \propto t^{0.3}
        \label{eqn:Lss_tscale} \\
    \end{equation}

    \subsubsection{Adiabatic Losses Light Curve}
    \label{subapp:adiabaticLC}

    After several dynamical times (Equation \ref{eqn:tdyn}) have passed 
    since shock breakout, the system should have relaxed into a state of free
    expansion, with $R \propto t$. In this state, we expect adiabatic evolution of the gas,
    \begin{eqnarray}
      V &\propto& R^3 \propto t^3 \\
      p &\propto& V^{-5/3} \; \Rightarrow\; \ugas \propto V^{-5/3} \propto t^{-5} ~ \;.
    \end{eqnarray}
    This produces a deline in the emissivity
    \begin{equation}
      j_\nu \propto \rho^{-1}~ \ugas^3 \propto t^{-12}
    \end{equation}
    so in this state of free expansion we have
    \begin{equation}
      L_\nu \propto j_\nu V \propto t^{-9} 
      \label{eqn:Ladiab} \; .
    \end{equation}
    This explains why the decline in the radio light curves is so
    steep after shock breakout.
    
    \subsubsection{Synchrotron Self-Absorption}
    \label{subapp:ssa}
    
    The relativistic electrons that create synchrotron emission can also
    absorb synchrotron radiation in a process known as synchrotron 
    self-absorption (SSA). 
    From \citet{RL1979}, the extinction coefficient for SSA is
    \begin{equation}
    	\alSSA = \frac{\sqrt{3}q^3}{8\pi m_e} 
        		\lp \frac{3q}{2\pi m^3c^5}\rp^{p/2} C_E (B\sin\theta)^{(p+2)/2}
                \Gamma\lp\frac{3p+2}{12}\rp \Gamma\lp\frac{3p+22}{12}\rp
                \nu^{-(p+4)/2}
    \end{equation}
	where $\Gamma$ is the gamma function and 
    $C_E$ is from $n(E)=C_E E^{-p} dE$ and is thus related to the 
    $C$ of Equation \ref{eqn:C} by $C_E = (mc^2)^{p-1} C$. We have 
    used $p=3$ in the second expression. As previously 
    mentioned, the $\sin\theta$ term is a pitch angle term that is attached
    to the magnetic field amplitude $B$ and thus carried through the derivation. 
    In our situation, we expect both the magnetic fields and the electron 
    velocities to be randomly oriented
    and therefore when using synchrotron equations from 
    \citet{RL1979}, we adopt the angle-averaged value
    $B\sin\theta = B\left< \sin \theta \right>_\theta = (2/\pi)B$.
    Thus with some substitution we can write this as 
    \begin{eqnarray}
    \alSSA &=& \frac{1}{64}  \lp \frac{6}{\pi}\rp^{(p+3)/2}
         \Gamma\lp \frac{3p+2}{12}\rp \Gamma\lp \frac{3p+22}{12}\rp
         \frac{\sigma_T c}{m_e} u_B C \nu_B^{(p-2)/2} 
         \nu^{-(p+4)/2} \\[10pt]
         &=& (4.74\times10^{-7}~\mathrm{cm^{-1}}) 
            \lp \frac{\epse}{0.1} \rp^2
            \lp \frac{\epsB}{0.1} \rp^{5/4}~\nonumber\\[8pt]
        &&\times
         	\lp \frac{\rho}{10^{-18}~\gccunit} \rp^{-1} 
            \lp \frac{\ugas}{10^2~\mathrm{erg~cm^{-3}}} \rp^{13/4}
	        \lp \frac{\nu}{\mathrm{GHz}} \rp^{-7/2}
        \label{eqn:alSSA}
    \end{eqnarray}
    where $\rho$ is the mass density of the shocked gas and $\ugas$ its
    internal energy density. 
    
    We can estimate the optical depth of a
    model by assuming $\alSSA$ is uniform in the shocked gas. 
    With $\mu\approx1$ and  $\rho = 4\rhocsm$ the energy density is
    \begin{equation}
    	\ugas \approx \frac{6\rhocsm}{m_p} k_B T \approx 0.5~ \rho_\mathrm{CSM,-18} T_9 ~
    \end{equation}
	so 
    \begin{equation}
		\alSSA \approx (1.6\times10^{-14}~ \mathrm{cm^{-1}})~
        		\rho_\mathrm{CSM,-18}^{2.25} T_{9}^{3.25} \nu_9^{-3.5}
    \end{equation}    
    where $\nu_9=\nu/\mathrm{GHz}$. 
    Assuming the shocked region has a width $\Delta R_s \sim 0.1R_c$ then
    \begin{eqnarray}
		\tau_\mathrm{SSA} &\sim& 0.1 \alSSA R_c \\
        					&\sim& 16~ R_{c,16}~\rho_\mathrm{CSM,-18}^{2.25}~
                            T_{9}^{3.25}~ \nu_9^{-3.5} ~.
        \label{eqn:tau_SSA}
    \end{eqnarray}
    The optical depth of a model varies with time because of these dependencies
    on temperature and radius, but we show the line of $\tau = 1$ in 
    Figure \ref{fig:paramspace} assuming $T_9 = 20$ and $\nu_9 = 4.9$ for 
    reference. 
    
	When fiducial models are optically thin such that the synchrotron luminosity is 
    set by the emissivity, their light curves obey scaling relations that reflect
    the similar hydrodynamic evolution of the models as described in \S~\ref{sec:shell_LCs}. 
    As the models become more optically thick, the light curve is set 
    by the source function $S_{\nu,\SSA}$, which from 
    Equations \ref{eqn:physj_p3} and \ref{eqn:alSSA} we calculate to be
    \begin{eqnarray}
    	S_{\nu,\SSA} = \frac{j_\nu}{\alSSA}
        &=& (2.8\times10^{-9}~\mathrm{erg~s^{-1}~cm^{-2}~Hz^{-1}~sr^{-1}})~\nonumber\\[8pt]
        &&\times        
        \lp \frac{u_B}{10~\mathrm{erg~cm^{-3}}} \rp^{-1/4}~ 
        \lp \frac{\nu}{\mathrm{GHz}} \rp^{-5/2}~.
    \end{eqnarray}

	For arbitrary optical depth, the luminosity of a one-dimensional 
    numerical model can be found via the formal solution of the radiative
    transfer equation using this source function.



\end{document}